\documentstyle[12pt,epsfig]{article}
\newlength{\dinwidth}
\newlength{\dinmargin}
\newcommand{\ba}{\begin{array}}
\newcommand{\ea}{\end{array}}
\newcommand{\beq}{\begin{equation}}
\newcommand{\eeq}{\end{equation}}
\newcommand{\bea}{\begin{eqnarray}}
\newcommand{\eea}{\end{eqnarray}}

\def\bce{\begin{center}}
\def\ece{\end{center}}

\def\nonu{\nonumber}

\def\al{\alpha}

\def\La{\Lambda}

\newcommand{\tr}{\mbox{Tr}}

\def\eps6{{\displaystyle \mathop{\epsilon}^{6}}{}}

\def\nab6{{\displaystyle \mathop{\nabla}^{6}}{}}
\def\to{\rightarrow}

\newcommand{\bean}{\begin{eqnarray*}}
\newcommand{\eean}{\end{eqnarray*}}

\begin{document}
\thispagestyle{empty} \addtocounter{page}{-1}
\begin{flushright}
TIT-HEP-501 \\
{\tt hep-th/0307190}\\
\end{flushright}

\vspace*{1.3cm} \centerline{\Large \bf Phases of} \vskip0.3cm
\centerline{ \Large \bf ${\cal N}=1$  $USp(2N_c)$ Gauge Theories with 
Flavors} \vskip0.3cm 

\vspace*{1.5cm}
\centerline{{\bf Changhyun Ahn}$^1$, {\bf  Bo Feng}$^2$  and 
{\bf Yutaka Ookouchi}$^3$}
\vspace*{1.0cm} \centerline{\it $^1$Department of Physics,
Kyungpook National University, Taegu 702-701, Korea}
\vspace*{0.2cm} \centerline{\it $^2$Institute for Advanced Study,
 Olden Lane, Princeton, NJ 08540, USA}
\vspace*{0.2cm} \centerline{\it $^3$Department of Physics,
 Tokyo Institute of
Technology, Tokyo 152-8511, Japan}

\vspace*{0.8cm} \centerline{\tt
ahn@knu.ac.kr,  \qquad fengb@ias.edu,}
\vspace{0.2cm}
\centerline{\tt
ookouchi@th.phys.titech.ac.jp}

 \vskip2cm

\centerline{\bf Abstract}
\vspace*{0.5cm}

We studied the phase structures 
of ${\cal N}=1$ supersymmetric $USp(2N_c)$ gauge theory
with $N_f$ flavors in the fundamental representation as we deformed
the ${\cal N}=2$ supersymmetric QCD by adding the superpotential 
for adjoint chiral scalar field. 
We determined
the most general factorization curves for various breaking patterns,
for example, the  two different breaking
patterns of quartic superpotential. 
We observed all kinds of smooth transitions for quartic superpotential. 
Finally we discuss the intriguing role of
$USp(0)$ in the phase structure and the possible
connection with  observations made recently 
in hep-th/0304271 (Aganagic, Intriligator, Vafa and Warner) 
and in hep-th/0307063 (Cachazo). 


\baselineskip=18pt
\newpage
\renewcommand{\theequation}{\arabic{section}\mbox{.}\arabic{equation}}

\section{Introduction and Summary}
\setcounter{equation}{0}

\indent

The ${\cal N}=1$ supersymmetric gauge theories in four dimensions
have  rich structures and the nonperturbative aspects can be characterized 
by 
the holomorphic effective superpotential which determines 
the quantum moduli space.  
A new recipe for the calculation of the effective superpotential
was proposed in \cite{dv1,dv2,dv3} through the  
free energies in the bosonic matrix model.
These matrix model analyses could be interpreted within purely field 
theoretic point of view 
\cite{cdsw}. 
A new kind of duality where one can transit several vacua 
with different broken gauge groups continuously and holomorphically by 
changing the parameters of the superpotential was given by \cite{csw}
(similar ideas have appeared in the earlier work  
\cite{tamar,ferrari,ferrariother}
and in the more recent one \cite{Casero}).
The extension of \cite{csw} to the ${\cal N}=1$ supersymmetric gauge
theories with the gauge group $SO(N_c)/USp(2N_c)$ 
was found in \cite{ao} where 
the phase
structures of these theories, the 
matrix model curve, and a generalized Konishi
anomaly equation were obtained. 
In \cite{bfhn}, 
by adding the flavors in the fundamental representation 
to the theory of \cite{csw}, the vacuum structures,  
an addition map, and multiplication map were developed.
Recently, 
the phase structures 
of ${\cal N}=1$ supersymmetric $SO(N_c)$ gauge theory
with $N_f$ flavors in the vector representation \cite{afo} were obtained. 

In this paper, 
we continue to study the phase structures 
of ${\cal N}=1$ supersymmetric $USp(2N_c)$ gauge theory
with $N_f$ flavors in the fundamental representation by deforming 
the ${\cal N}=2$ supersymmetric QCD with the superpotential of
arbitrary polynomial for the adjoint chiral scalar field, by applying the 
methods in \cite{csw,ao,bfhn,afo}. These kinds of study were initiated
in \cite{csw}, in which a kind of new
duality was found. 
This paper is a generalization of \cite{ao}
to the $USp(2N_c)$ with flavors. We found that with flavors, the phase
structure is richer and that more interesting dualities show up.     
We refer to \cite{afo}  for some relevant papers  
on the recent works, along the line  of 
\cite{dv1,dv2,dv3}.

In section 2,
we describe the classical moduli space of
${\cal N}=2$ SQCD deformed to ${\cal N}=1$ theory by adding the 
superpotential $W(\Phi)$ (\ref{treesup}). The gauge group $USp(2N_c)$ will
break to $USp(2N_0)\times \prod_{j=1}^n U(N_j)$ with $2N_0+\sum_{j=1}^n 
2N_j=2N_c$
by choosing the adjoint chiral field $\Phi$ to be the root of
$W'(x)$  or $\pm i m_i$, the mass parameters. To obtain a pure Coulomb branch
where no any factor $U(N_j)$ is higgsed, we restrict ourselves to
the case $W^{\prime}(\pm i m_i)=0$. For each factor with some effective
massless flavors, 
there exists a rich structure
of Higgs branches, characterized by an integer $r_i$, 
that meets the Coulomb
branch along the submanifold. 

In section 3.1,
we discuss the quantum moduli space of $USp(2N)$ by both the
weak and strong coupling analyses.
When the difference between the roots of $W^{\prime}(x)$ is much 
larger than
${\cal N}=2$ dynamical scale $\La$ (in the weak coupling
region), the adjoint scalar field $\Phi$ can be integrated out, giving
a low energy effective ${\cal N}=1$ superpotential. Under this condition,
the higher order terms except for the quadratic piece 
in the superpotential  (\ref{treesup}) can be ignored. 
Then the effective superpotential 
consists of the classical part plus nonperturbative part. We summarize the
quantum theory with  mass deformation for $\Phi$ in the weak coupling
analysis. There exist 
two groups of solutions, i.e., {\it Chebyshev} vacua and 
{\it Special} vacua,
according to the unbroken flavor symmetry, a meson-like matrix $M$, 
and  various phases of 
vacua. 
 At  scales below
the  ${\cal N}=1$ scale $\La_1$ (when the roots of
$W^{\prime}(x)$ are almost the same), 
 strong coupling analysis is relevant.     
We need to  determine the special points where some number of 
magnetic monopoles (mutually local or non local)  
become massless, on the 
submanifold of the Coulomb branch of  ${\cal N}=2$ $USp(2N)$ which
is not lifted by the ${\cal N}=1$ deformation. The conditions for
these special points are translated into a
particular factorization form
of the corresponding Seiberg-Witten curve. 
We discuss some aspects of these curves
at the Chebyshev branch or the Special branch, in particular, the 
power of factor $t=x^2$ and the number of single roots.  

In  section 3.2, 
combining the quantum moduli space of $USp(2N)$ group with the
quantum moduli space of $U(N)$ group studied in \cite{bfhn,afo},
we give the most general factorization 
curves with the proper number of single roots and double roots
for various symmetry breaking patterns, which generalize the
results in \cite{fo1,cv,feng1,ookouchi}.
From the point of view
of the geometry, these various breaking patterns correspond to
the various distributions of wrapping D5-branes among the 
roots of $W'(x)$. In this subsection, {\it we also describe 
the mysterious role
of $USp(0)$ in the phase structure }.

In section 4, we study the quartic tree level superpotential 
with {\it massive} flavors for $USp(4)$ and $USp(6)$ gauge groups.
There exist two breaking patterns 
$USp(2N_c)\to USp(2N_0)\times U(N_1)$ where $N_0+N_1=N_c,~~N_0\geq 0$ 
and 
$USp(2N_c)\to USp(2N_c)$. 
Depending on the properties of various factors, the factorization 
problems lead to  interesting smooth transitions 
among these two breaking patterns:
  $USp(2N_c)\leftrightarrow USp(2M_0)\times U(M_1)$,
$USp(2N_0)\times U(N_1)\leftrightarrow USp(2M_0)\times U(M_1)$
and $USp(N_0) \times U(N_1) 
\leftrightarrow USp(2M_0)\times U(M_1) \leftrightarrow 
USp(2L_0)\times U(L_1) $. The phase structures for various product 
gauge groups are summarized in the Tables in section 4 and 
section 5. The addition map application helps us to derive the phases
from known vacua, without computing the details.

In section 5,
 we move to the {\it massless} flavors
with quartic deformed superpotential for $USp(2N_c)$ 
where $N_c=2$ and $3$. In this case, 
at the IR limit the $\widehat{USp(2N_0)}$ factor has massless 
flavors instead of $U(N_i)$ factor. Because of this difference, new features
arise. 
For the
smooth transition 
$\widehat{USp(2N_0)} \times U(N_1)\leftrightarrow 
\widehat{USp(2M_0)} \times U(M_1)$
in the Special branch,
we have $M_0=(N_f-N_0-2)$ ($M_0 < N_0$ and  
$ 2M_0+2 \leq N_f < 2N_0+2$), 
which is the relationship between
$\widehat{USp(2N_0)}$ and $\widehat{USp(2M_0)}$ 
to be the {\sl Seiberg dual} pair. 
In fact, the smooth transition in this case may be rooted in
the Seiberg duality. 

In section 6, we use the Brane setup to understand the mysterious role
of $USp(0)$ and recent observations made in 
\cite{vafa, Freddy}. 

In Appendix A,
by using the ${\cal N}=2$ curve together with monopole constraints we are 
interested in and applying the contour integral formula, we derive  
the matrix model curve (\ref{matrixcurve}) 
for deformed superpotential with an arbitrary degree
and the relationship (\ref{spnewmatrix}) (or the most general
expression 
between the matrix model curve 
and the deformed superpotential $W'(x)$). 
Using these results,
we have also checked  the generalized Konishi anomaly equation 
for our gauge theory with flavors (\ref{Konequation}).

In Appendix B,
in order to understand the vacua of different theories, 
we discuss both the addition map and multiplication map. 
The addition map with  flavors relates
the vacua of $USp(2N_c)$ gauge theory with $N_f$ flavors 
in the $r$-th branch 
to those of $USp(2N_c^{\prime})$ gauge theory with $N_f^{\prime}$ flavors 
in the $r^{\prime}$-branch. 
This phenomenon is the  same as the one in the 
$U(N_c)$ gauge theory with flavors. 
For the multiplication map, we present the most general form.
Through this multiplication map, one  obtains the unknown 
factorization of the gauge group with higher rank 
from the known factorization of the gauge 
group with lower rank. 
One can construct a multiplication map from $USp(2N_c)$
with $2l$ {\it massive} flavors to $USp(K(2N_c+2)-2)$ with $2Kl$ 
{\it massive} flavors
where $K$ is a positive integer.
In this derivation, the properties of Chebyshev 
polynomials are used.


\section{The classical moduli space  of $USp(2N_c)$ supersymmetric 
QCD}
\setcounter{equation}{0}

\indent

Although the classical picture will be changed by quantum corrections, 
in certain situations, the classical analysis can be used to find out some 
information that is also valid in the full quantum theory.


Let us discuss ${\cal N}=1$ supersymmetric 
$USp(2N_c)$ gauge theory with $N_f$  flavors of quarks $Q_a^i$ 
where $a=1 \cdots, 2N_c$ and $i=1, \cdots, 2N_f$ 
in the fundamental representation. 
The tree level superpotential of the theory can be obtained from 
${\cal N}=2$ SQCD by including the arbitrary polynomial of the adjoint
scalar $\Phi$ belonging to the ${\cal N} =2$ vector multiplet:
\bea
\label{USp-W-1}
W_{tree}(\Phi,Q)=\sqrt{2} Q_a^i \Phi^{a}_b J^{bc} Q_c^i +
\sqrt{2} m_{ij} Q_a^i Q_b^j
J^{ab}
+\sum_{s=1}^{k+1} g_{2s} u_{2s}
\eea 
where the first two terms come from the ${\cal N}=2$ theory and the third
term, $W(\Phi)$, is a small perturbation of
${\cal N}=2$ $USp(2N_c)$ gauge theory 
\cite{as,aps,kitao,tera96,ahn98,tera97,Konishi-SU,eot,feng1} 
\begin{eqnarray}
W(\Phi)=\sum_{s=1}^{k+1}\frac{g_{2s}}{2s}\mbox{Tr}\Phi^{2s}\equiv
\sum_{s=1}^{k} g_{2s} u_{2s}, \qquad u_{2s} \equiv \frac{1}{2s}
\mbox{Tr} \Phi^{2s}
\label{treesup}
\end{eqnarray}
where $\Phi$ 
plays the role of
a deformation breaking ${\cal N}=2$ supersymmetry to ${\cal N}=1$ 
supersymmetry. 
The $J_{ab}$ is the symplectic metric  and $m_{ij}$ is a quark mass that 
together are 
represented as
\bea
J_{ab}=\left(\begin{array}{cc} 0 & 1 \\ -1 & 0 \end{array} \right)
\otimes {\bf I}_{N_c\times N_c},~~~
m=\left(\begin{array}{cc} 0 & -1 \\ 1 & 0 \end{array} \right)
\otimes \mbox{diag}(m_1, \cdots, m_{N_f})
\nonu
\eea
where the symplectic metric $J_{ab}$ is used to raise or lower the 
$USp(2N_c)$
color indices. The ${\cal N}=2$ 
theory is asymptotically free for $N_f<2N_c+2$ (generating
a strong coupling scale $\La$), conformal for
$N_f=2N_c+2$ and IR free for $N_f>2N_c+2$. 

The classical vacuum structure, the zeroes of the scalar potential,
can be obtained by solving 
D-terms and F-terms.
We summarize the following results from
the mechanism of adjoint vevs as follows:

1)
The eigenvalues of $\Phi$, $ \pm \phi_j$ can only be  the roots of $W'( x)$
or $\pm im_j$; thus, the  gauge group $USp(2N_c)$  with $N_f$  flavors 
is broken to 
the product of blocks with or without the effective massless flavors.
Among these blocks, at most one
block is $USp(\mu_{n+1})$ and others, $U(\mu_i)$ where $i=1, 2, \cdots, n$. 

2)
If $\phi_i=\pm i m_i$ but $W'(\pm im_i)\neq 0$, the 
corresponding gauge symmetry of that block will be completely higgsed.
Due to this fact, in the following discussions we will restrict ourselves
to the case of $W'(\pm im)= 0$ where there exist richer structures.

3)
For each block, if it has the ``effective'' massless flavors, the
vacua are classified by an integer $r$. 

\section{The quantum moduli space of $USp(2N_c)$ supersymmetric 
QCD }
\setcounter{equation}{0}

\indent

According to the previous section, 
the gauge group $USp(2N_c)$ is broken to the product of 
various gauge group $U(N_i)$ with at most one single $USp(2N)$.
In order to understand the quantum moduli space of $USp(2N_c)$
supersymmetric QCD, we need to study the quantum moduli for each 
factor gauge group.
In this section, we focus on the quantum moduli space of $USp(2N)$
gauge group with {\it massless} flavors discussed in 
\cite{aps,Konishi-SU} because
the quantum moduli of $U(N_i)$ with effective {\it massless}  flavors
were already studied in \cite{aps1,Konishi-SU,bfhn}.

\subsection{The quantum theory of $USp(2N)$ supersymmetric QCD}

\indent

The quantum theory of   $USp(2N)$ with mass deformation 
$\frac{1}{2} \mu\tr\Phi^2$
has been described in \cite{aps,Konishi-SU} 
for both weak and strong coupling 
analyses. 
We will summarize the main results.

\subsubsection{The weak coupling analysis}

\indent

When the mass $\mu$ for the adjoint scalar $\Phi$ is larger than 
the ${\cal N}=2$ dynamical scale $\La$ 
(that is, $\mu \gg \Lambda$), we integrate out $\Phi$ first and 
by resubstituting the $\Phi$ into the superpotential (\ref{USp-W-1})
we get
$-{1\over 8 \mu} \tr(MJMJ)$ where $M^{ij}=Q_a^i J^{ab} Q_b^j=-M^{ji}$
\footnote{Notice that for $USp(2N)$ gauge group, there is no
baryonic-like gauge invariant operator. 
The invariant tensor breaks up into sums of products of the 
$J^{ab}$ and baryons break up into mesons \cite{ip}. 
In this subsection, we consider only massless 
flavors (the quark mass $m$ vanishes).}
are the meson-like composite superfields. However, depending on 
the number of flavors, various terms can be added to the superpotential
by quantum effects. 
To find   ${\cal N}=1$ vacua, the effective superpotential
should be minimized.  We can study the 
vacuum structure, the number of vacua, and  global symmetry
breakings according to the range of the number of flavors $N_f$.

We summarize two groups of solutions as follows:

1) The first group exists
for arbitrary number of flavors with broken flavor symmetry 
$SO(2N_f)\rightarrow
U(N_f)$ and the counting of vacua is $(2N+2-N_f)$. 
It corresponds to the 
{\it Chebyshev
point} (which will be discussed soon) 
where its position in the ${\cal N}=2$ 
moduli space is located by the roots of a Chebyshev polynomial. 
The light degrees of freedom are mutually nonlocal and the theory flows 
to an 
interacting ${\cal N}=2$ superconformal theory.  
The dynamic flavor symmetry breaks into $U(N_f)$.

2) The second group exists only when
$N_f\geq N+2$ with unbroken flavor symmetry $SO(2N_f)$. It  corresponds
to the ``baryonic-like'' root. In this  
{\it Special point}, the gauge symmetry is enhanced to 
$USp(2\widetilde{N}) 
\times U(1)^{N-\widetilde{N}}$ where $\widetilde{N}=N_f-N-2$ 
and the full $SO(2N_f)$ global symmetry remains unchanged
since there are no meson condensates and no dynamic symmetry breaking 
occurs. 
These vacua are in the  free magnetic phase.   

\subsubsection{The strong coupling analysis}

The strong coupling analysis has been done in \cite{aps,Konishi-SU}. 
Let us recall the curve 
of $USp(2N)$ \cite{as} first
\bea
ty^2= \left[t\prod_{j=1}^{N} (t-\phi_j^2)+ 2 \Lambda^{2N+2-N_f}
\prod_{k=1}^{N_f} m_k \right]^2- (-1)^{N_f} 4  \Lambda^{2(2N+2-N_f)}
\prod_{k=1}^{N_f} (t-m_k^2)
\nonu
\eea
where we have used a relation $t=x^2$. 
When $2N+2-N_f=0$ corresponding to the conformal theory,  we should replace
$2 \Lambda^{2N+2-N_f}$ by  the moduli form
\bea
g(\tau)={{\cal \theta}_2^4 \over {\cal \theta}_3^4+{\cal 
\theta}_4^4}.
\nonu
\eea
For our analysis where all $m_k=0$, the curve is simplified to be
\bea
\label{USp-curve-massless}
ty^2= \left[t\prod_{j=1}^{N} (t-\phi_j^2) \right]^2-
4  \Lambda^{2(2N+2-N_f)}
t^{N_f}.
\eea
If $r$ $\phi_j$'s  vanish and the remaining $(N-r)$'s do not, 
we can factorize the curve (\ref{USp-curve-massless})
as 
\bea
\label{USp-factor-r}
ty^2= t^{2(r+1)} \left[\prod_{j=1}^{N-r} (t-\phi_j^2)^2-4 
\Lambda^{2(2N+2-N_f)} t^{N_f-2(r+1)} \right].
\eea 
The analysis done in \cite{aps,Konishi-SU} implies
that only when 
\bea
r=\widetilde{N}=N_f-N-2
\nonu 
\eea 
and 
\bea
r=N_f/2-1, \qquad 
\mbox{for} \;\;
N_f \;\; \mbox{even}
\nonu
\eea 
or 
\bea
r=(N_f-1)/2, \qquad
\mbox{for} \;\; N_f \;\; \mbox{odd}, 
\nonu
\eea 
the curve (\ref{USp-factor-r})
\footnote{
When $N_f$ is even,
since $\phi_1=\phi_2= \cdots = \phi_{r=N_f/2-1}$=0, the curve becomes 
$ty^2= t^{N_f} \left[ \prod_{j=1}^{N+1-N_f/2} (t-\phi_j^2)^2 -
4  \Lambda^{2(2N+2-N_f)} \right]$.
When $N_f$ is odd,
since $\phi_1=\phi_2= \cdots = \phi_{r=(N_f-1)/2}$=0, the curve becomes 
$ty^2= t^{N_f} \left[t\prod_{j=1}^{N-(N_f-1)/2} (t-\phi_j^2)^2 -
4  \Lambda^{2(2N+2-N_f)} \right]$.}
gives the vacua which are {\it not} lifted by the mass deformation 
$\frac{1}{2} \mu \tr \Phi^2$.

The basic idea of the above results is as follows. For general $r$
the low energy theory is of the trivial conformal theory,
i.e., the class 1 in the classification of \cite{Hori}. For these theories,
to have unlifted vacua, we must have a sufficient number of mutually local 
massless monopoles, i.e., enough double roots in the SW-curve. 
For the $USp(2N)$ gauge group, it is necessary that $ty^2$ is
a total square form which can happen if and only if 
$r=\widetilde{N}=(N_f-N-2)$ by noticing that
\bea
ty^2 =  t^{2(\widetilde{N}+1)} \left(P^2_{N-\widetilde{N}}(t)-
4\Lambda^{2(N-\widetilde{N})} t^{N-\widetilde{N}} \right) 
 =  t^{2(\widetilde{N}+1)}  \left(t^{N-\widetilde{N}}-
\Lambda^{2(N-\widetilde{N})} \right)^2
\nonu
\eea
corresponding to $(9.137)$ of \cite{Konishi-SU}
when we choose $P_{N-\widetilde{N}}(t)=t^{N-\widetilde{N}}+
\Lambda^{2(N-\widetilde{N})}$ \footnote{In this case, one takes 
$(\phi_1^2, \phi_2^2, \cdots, \phi_{N-\widetilde{N}}^2)=\La^2 (\omega, 
\omega^3, \cdots, \omega^{2(N-\widetilde{N})-1})$ where $\omega=e^{\pi i/(N-
\widetilde{N})}$. Then the expression of $P_{N-\widetilde{N}}(t)=\prod_{j=1}^
{N-\widetilde{N}} (t -\phi_j)$ can be written as $t^{N-\widetilde{N}}+
\Lambda^{2(N-\widetilde{N})}$. The locations of double zeros of the factor in 
the parenthesis are at $t=\La^2 \omega^2, \La^2 \omega^4, \cdots, 
\La^2 \omega^{2(N-\widetilde{N})}=\La^2$. }. It looks like the baryonic root 
$r=N_f-N$ of the $U(N)$ gauge group, so we will call it the
baryonic root of $USp(2N)$. As shown in \cite{Konishi-SU},
after perturbation by mass term of flavors,  $r=\widetilde{N}$
will give the second group ({\it Special Point}) 
of solutions discussed in the previous
subsection.

Except for the above trivial conformal theory, there exists a special
case in which  the low energy effective theory is of non-trivial
conformal theory. It is when $r=N_f/2-1$ for
$N_f$ even or $r=(N_f-1)/2$ for $N_f$ odd, which is the class 2 or 3
in the classification of \cite{Hori} respectively,  and is called 
the {\it Chebyshev point}. To see it, assuming that $N_f$ is even, the curve
becomes $ty^2= t^{N_f}[ P^2_{N+1-N_f/2}(t)-4\Lambda^{4(N+1-N_f/2)}]$.
When we take \footnote{The reason we define $\widetilde{t}=
{\sqrt{t} \over 2\eta \Lambda}$ is because $t$ has dimension two.
We use a useful relation ${\cal T}_K^2(x)-1 = (x^2-1) 
{\cal U}_{K-1}^2(x)$.}
\bea
P_{N+1-N_f/2}(t)=2(\eta \Lambda)^{2(N+1-N_f/2)} {\cal T}_{2(N+1-N_f/2)}
\left({\sqrt{t} \over 2\eta \Lambda} \right)
\nonu
\eea
with $\eta^{4(N+1-N_f/2)}=1$, the curve becomes
\bea
ty^2 & = &  t^{N_f} 4(\eta \Lambda)^{4(N+1-N_f/2)}
\left[{\cal T}^2_{2(N+1-N_f/2)}
 \left({\sqrt{t} \over 2\eta \Lambda} \right)-1 \right] \nonu \\
& = &  t^{N_f} 4(\eta \Lambda)^{4(N+1-N_f/2)} 
\left[ \left({\sqrt{t} 
\over 2\eta \Lambda}\right)^2-1 \right]
{\cal U}_{2(N+1-N_f/2)-1}^2
\left({\sqrt{t} \over 2\eta \Lambda} \right). 
\nonu
\eea
Notice that although $\eta^{4(N+1-N_f/2)}=1$, the characteristic function 
$P_{N+1-N_f/2}(t)$
is a function of $t$ and $\eta^2$ because only the even power term appears
in a first kind of Chebyshev polynomial ${\cal T}_{2(N+1-N_f/2)}
 \left({\sqrt{t} \over 2\eta \Lambda} \right)$ 
with even order $2(N+1-N_f/2)$; thus, we have $(2N+2-N_f)$ solutions as
given in the weak coupling analysis. Additionally, note that 
except for a factor $t^{N_f}$, unlike the case of the total square form
of $r=\widetilde{N}$, there is a  single root
at $t=0$ and 
at $t=4\eta^2\Lambda^2$ from the factor 
$\left[ \left({\sqrt{t} 
\over 2\eta \Lambda}\right)^2-1 \right]
{\cal U}_{2(N+1-N_f/2)-1}^2
\left({\sqrt{t} \over 2\eta \Lambda} \right)$ since the second kind of
Chebyshev polynomial ${\cal U}_K(x)$ with an 
odd degree $K$ has a common factor 
$x$ and so  
there exists  a factor $t^{N_f+1}$ which is odd for this case.
Moreover, there are double zeros at $(N-N_f/2)$ different locations of $t$
from the structure of ${\cal U}_{2(N+1-N_f/2)-1}^2
\left({\sqrt{t} \over 2\eta \Lambda} \right)$.
A similar discussion can be carried
out for $N_f$ odd. It turns out that there are $N_f$ (which is odd) 
zeros at $t=0$, 
a single zero at $t=4\eta^2\Lambda^2$ and double zeros at $N-(N_f-1)/2$ 
different locations of $t$.

\subsection{The factorized form of a hyperelliptic curve} 

\indent

Now combining the proper factorization form of $USp(2N)$ curve 
with the massless flavors in the previous subsection with the  
$U(N_i)$ curve with flavors in \cite{bfhn}, 
the general form of the curve can be obtained. First we 
need to have the proper prefactor (like $t^p$ for $USp(2N)$ part and 
$(t-m^2)^{2r}$ for $U(N_i)$ part at the $r$-th branch). 
After factorizing out these prefactors, then we require the 
remaining curve to have a proper number of double roots 
and single roots, which will fix finally the form of factorization. 

For every $U(N_i)$ 
factor at the non-baryonic branch, there exist {\it two} single roots 
while there is {\it no} single root at the baryonic branch. 
For the possible $USp(2N)$ factor, when it is at 
the Chebyshev branch, there are {\it two} single roots where one 
of them is at the origin, i.e., we have a factor $t$ for that single 
root. However, when it is at the Special branch, there exists 
{\it no} single root for the block $USp(2N)$. Adding 
the single roots together for all the blocks, the 
final number of the single roots can be found. 
Additionally  notice that the Special branch 
(or the baryonic branch) will have {\it one} more double root 
than the Chebyshev branch (or the non-baryonic branch) 
in the factorization form.

However, there is one important point which differentiates the 
$U(N)$ and $SO(N)$ gauge groups; {\it although the $USp(0)$ does 
not have the dynamical meaning, it does effect the form of 
factorization of the Seiberg-Witten curve.} In other words, 
by including the $USp(0)$ factor group,
the breaking pattern $USp(2N_c)\to U(N_c)$ should be really 
written as,  
\bea 
USp(2N_c)\to U(N_c) \times USp(0). 
\nonu
\eea 
The specialty of $USp(0)$ can be 
seen from the strong coupling analysis. Firstly, 
the number of vacua of 
$USp(2N)$ at 
the Chebyshev branch is $(2N+2-N_f)$ for $N_f\geq 1$ or 
$(N+1)$ for $N_f=0$. Setting $N=0$ we find one for 
$N_f=0,1$ and zero for $N_f\geq 2$. Secondly, the power of 
a factor $t$ at the Special branch is $(N_f-N-1)$, which is a 
positive number if $N_f\geq 2$. Combining these two 
observations, we get the following conclusion: {\it $USp(0)$ has 
the Chebyshev branch with a factor $t$ for $N_f=0,1$ and 
the Special branch with a factor 
$t^2$ for $N_f\geq 2$.}

Now we can discuss the general factorization form of the Seiberg-Witten 
curve by using the following simplest nontrivial examples 
in which $USp(2N)$ gauge group is broken to the following two cases: 
\bean 
USp(2N_c)\to USp(2N_0)\times U(N_1), ~~~~(N_0+N_1=N_c,~~N_0\geq 0) 
\eean 
and 
\bean 
\qquad USp(2N_c)\to USp(2N_c). 
\eean 
(and the generalization to multiple blocks 
will be straightforward.) 

For the broken pattern 
$USp(2N_c)\to USp(2N_0)\times U(N_1)$, by counting the number of the 
single 
roots, the following four possible curves are obtained: 
\bea 
t y^2 & = & t F_3(t) H^2_{N_c-2}(t), \label{example-1}  \\ 
t y^2 & = & F_2(t) H^2_{N_c-1}(t), \label{example-2} \\ 
t y^2 & = & t F_1(t) H^2_{N_c-1}(t), \label{example-3} \\ 
t y^2 & = & H^2_{N_c}(t). \label{example-4} 
\eea 
The curve (\ref{example-1}) is 
for $USp(2N_0)$ at the Chebyshev branch, and $U(N_1)$ at the 
non-baryonic branch. First we should have 
four single roots. 
Secondly, because $USp(2N_0)$ is at 
the Chebyshev branch, one of the four single roots must be 
at the origin $t=0$ and finally one gets $F_4(t)=t F_3(t)$. 
The curve (\ref{example-2}) is 
for $USp(2N_0)$ at the Special branch and $U(N_1)$ at the 
non-baryonic branch. The factor $F_2(t)$ with two single 
roots will record the information of $U(1)\subset U(N_1)$. 
The curve (\ref{example-3}) is 
for $USp(2N_0)$ at the Chebyshev branch and $U(N_1)$ at the 
baryonic branch. Since in this case, $USp(2N_0)$ is at the Chebyshev branch 
 one gets $F_2(t)=tF_1(t)$. Finally, 
the curve (\ref{example-4}) is 
for $USp(2N_0)$ at the Special branch and $U(N_1)$ at the 
baryonic branch, where no single root is required. 
Notice that the function $H_{p}(t)$ will have a proper 
number of $(t-m^2)$ or $t$ to count the prefactor 
for various branches.

For the unbroken pattern $USp(2N_c)\to USp(2N_c)$, one gets 
\bea 
\label{example-7} 
ty^2 & = & t F_1(t) H^2_{N_c-1}(t), \\ 
ty^2 & = & H^2_{N_c}(t), \label{example-8} 
\eea 
where the curve (\ref{example-7}) is for $USp(2N_c)$ at the 
Chebyshev branch, and the curve (\ref{example-8}) is  
for $USp(2N_c)$ at the Special branch. The function $H_{p}(t)$ will 
have a proper number of factor $t$ to count the 
prefactor required by the Special or Chebyshev branch. It is 
noteworthy that although both  curve (\ref{example-3}) 
and  curve (\ref{example-7}) look the same, they can be distinguished 
by  factor $H_{p}(t)$, where different powers of $t$ and $(t-m^2)$ 
can arise. \footnote{As we will see in the examples later, all the 
factorizations (\ref{example-1})-(\ref{example-8}) 
except (\ref{example-4}) appear in our study.  } 

For $U(N_i)$ with $M_i$ flavors, the counting of vacua has been 
given in \cite{bfhn,afo} as 
\begin{eqnarray} 
\mbox{The number of vacua}&=&\left\{ 
\begin{array}{lll} 
2N_i-M_i & r < M_i/2, & M_i \leq N_i, \\ 
N_i-M_i/2 & r = M_i/2, & M_i \leq N_i, \\ 
2N_i-M_i & r \geq M_i-N_i, & M_i \geq N_i+1, \\ 
N_i- r & r < M_i-N_i, & M_i \geq N_i+1, \\ 
1 & r =N_i -1(\mbox{nonbaryonic}), & \\ 
1 & r=M_i-N_i, N_i(\mbox{baryonic}). 
\end{array} \right. 
\label{vacua} 
\end{eqnarray} 
For pure $USp(2N)$ the counting of vacua is $(N+1)$. 
For $USp(2N)$ with $M\geq 1$ flavors, there are $(2N-M+2)$ vacua 
from the Chebyshev branch (which means that there is 
no Chebyshev vacua if $M\geq 2N+2)$) and one vacuum from the Special 
branch if $M\geq N+2$. 
The total number of vacua is the product 
of the number of vacua of various blocks.

\section{Quartic superpotential with massive flavors}
\setcounter{equation}{0}

\indent

The general curve of $USp(2N)$ gauge theory with $M$ flavors 
\cite{as,aps,bfh}
should be 
\bea
ty^2= \left[ t P_N(t) + 2 \La^{2N+2-M} \prod_{j=1}^M m_j \right]^2
-(-1)^M   4 \La^{2(2N+2-M)}\prod_{j=1}^M (t-m_j^2).
\nonu
\eea
It is noteworthy that the factor 
$(-1)^M$  will be crucial for later 
analysis. The presence of the factor $(-1)^M$ makes sure 
that there is always a factor $t$
in the curve because the $t$-independent two 
terms on the right hand side
 exactly cancel each other.

There is a difference of the counting for the vacua between the 
gauge groups $U(N)$ and $USp(2N)$. For pure $U(N)$ gauge group, 
the number of vacua
are $N$. However, with $N_f$ flavors it becomes $(2N-N_f)$, which shows 
very 
different behavior. For a pure $USp(2N)$ gauge group, 
the number of vacua is 
$(N+1)$ while with $N_f$
flavors, it is given by  $(2N+2-N_f)$.
The new aspect for the phase on the breaking 
$USp(2N_c)\to \widehat{U(N_c)}$
is that it is, in fact, the breaking pattern 
$USp(2N_c)\to \widehat{U(N_c)} \times USp(0)$, 
as we have described in section 3. 
For pure $USp(0)$, the number of
vacua is given by $(N+1)=1$ according to the above analysis, 
so it is {\it not} zero. \footnote{For $USp(0)$ with
$N_f$ flavors, the number of vacua for the
Chebyshev branch is $(2N+2-N_f)=1$ only if $N_f=1$. 
For the cases
$N_f\geq 2$, there is {\it no} Chebyshev branch for the factor 
$USp(0)$ due to the negativity of the vacua.
However, by the relation $(N_f-2-N)=N_f-2$, when $N_f\geq 2$,
 we have, instead, 
{\it Special branch} with a factor $t^2$ from the discussion of
section 3.}

Let us recall that the baryonic branch for $\widehat{U(N_1)}$ 
factor can be characterized by 
\bean
 r =N_1, \qquad \mbox{or} \qquad r =N_f-N_1
\eean
and that the index $r$ satisfies
\bean
0 \leq r \leq [N_f/2].
\eean
The number of flavors 
$N_f$ is restricted to $N_f < 2N_c+2$ for asymptotically free theory.
For a given number of colors $N_c$, both the number of flavors $N_f$
and index $r$ are fixed.
Now we are ready to consider explicit examples for $USp(4)$ and
$USp(6)$ gauge theories with massive flavors.

\subsection{ $USp(4)$ case} 

\subsubsection{$N_f=0$}

The curve is given by
\bea
t y^2= \left[ t (t^2-s_1t +s_2)+ 2\La^6 \right]^2-4 \La^{12}
\nonu
\eea
by 
denoting the characteristic function as $P_2(t)=t^2-s_1t +s_2$.

$\bullet$ {\bf Non-baryonic $r=0$ branch}

We need to have the following factorization 
\bea
t y^2= t F_3(t) H^2_1(t) = t (t^3 +a t^2+ bt+c) (t+d)^2.
\label{flavorless4}
\eea
The reason is that for $N_f=0$, there is { \it no} Special branch for
the $USp(2N_0)$ part and the power of a factor $t$ of the 
Chebyshev branch becomes
$N_f+1=1$ which  coincides with the one on the right hand side of 
the above curve. There are two kinds of solutions. The first kind of
solutions is given by
\bean
s_1=-2d,~~~s_2=d^2,~~~a=2d,~~~b=d^2,~~~~c=4 \La^6.
\nonu
\eean
The classical limit $\La \rightarrow 0$ implies that the 
curve reduces to  $t^2 (t+d)^4$, so it gives the breaking
pattern $USp(4)\to USp(0) \times 
U(2)$ because the characteristic function becomes
$P_2(t)=(t+d)^2$. The second kind of solution is summarized as 
follows:
\bean
s_1& = & -2d-{4 \La^6\over d^2},~~~s_2= d^2+{6 \La^6\over d},~~~
a=2d+{8 \La^6\over d^2},\\
b& = & d^2+{16\La^6\over d}+{16\La^{12}\over
d^2},~~~c=4 \left(\La^6+{8\La^{12}\over d^3} \right).
\nonu
\eean
There are two limits we can take: (1) If $\La\to 0$, but 
$d\to \mbox{constant}$,
the symmetry breaking patterns occur when $USp(4)\to U(2)\times USp(0)$; 
(2) If $\La\to 0$, $d\to 0$, but $\La^6/d^2 \to
\mbox{constant}$, we have a symmetry breaking 
$USp(4)\to U(1)\times USp(2)$ where the characteristic function 
becomes $P_2(t)=t(t+4\La^6/d^2)$. Thus, we see there exists a
smooth transition between $ U(2) \times USp(0)$ 
and $ U(1)\times USp(2)$.

Now let us count the number of vacua. For the non-degenerated case, we
require 
\bean
F_3(t)= W'(x)^2+{\cal O}(t)= t(t-m^2)^2+{\cal O}(t)
\Rightarrow -2m^2= a
\nonu
\eean
where the quadratic part of $t$ in $F_3(t)$ is 
identified with the one in
$W'(x)^2$. The
first kind of solution gives one vacuum. One special feature of
this case is that the curve will have a factor $(t+d)^2=(t-m^2)^2$ where
we put $d=-m^2$. However,
this does not mean it is at the $r=1$ branch, because there is no
$r=1$ branch. 
The second kind of solution gives three vacua. One of
them gives the breaking pattern  $USp(4)\to U(2)\times USp(0)$ 
(therefore,  there exist a total of two vacua which can also be seen 
from  counting in the pure $U(2)$ gauge theory) and two others give
$USp(4)\to U(1)\times USp(2)$ where there exist two roots for $d$:
$-2m^2=a=8\La^6/d^2$ where we know the $USp(2)$ has the number of vacua
by $(N_0+1)=2$.

We need to determine the $USp(4)\to USp(4)$. To achieve this, we require
an extra double root in the curve where $F_3(t)=F_1(t) K^2_1(t)$. 
Putting these
extra conditions into the three solutions we have found, all these 
cases give a pattern 
$USp(4)\to USp(4)$ at the classical limit. 
This matches the counting of vacua
$(N+1)=3$ in this case.

\subsubsection{$N_f=1$ }

The curve can be written as
\bea
t y^2= \left[ t (t^2-s_1t +s_2)+ 2\La^5 m \right]^2+4 \La^{10}(t-m^2).
\nonu
\eea
Note the sign of the second term in the right hand side.

$\bullet$ {\bf Baryonic and Non-baryonic $r=0$ branches}
 
From (\ref{example-1}),
first we require that the factorization should be
\bean
t y^2= t F_3(t) H^2_1(t) = t (t^3 +a t^2+ bt+c) (t+d)^2.
\nonu
\eean
There are two kinds of solutions where 
\bean
s_1 & = & - \frac{2\left( d^3 + {\La}^5 m
          \right)  \pm \frac{{\La}^5
          \left( d + 2 m^2 \right) }{{\sqrt{d + m^2}}}}{d^
       2}, \\
s_2 & = & \frac{d^3 + 4{\La}^5 m \pm  
     \frac{{\La}^5
        \left( 3 d + 4 m^2 \right) }{{\sqrt{d + m^2}}}}{d}.
\nonu
\eean
There are the following limits: (1) as 
$\La\to 0$, but $d\to \mbox{constant}$,
we have $USp(4) \to \widehat{U(2)}\times USp(0)$ where 
 the characteristic function behaves as
$P_2(t)=(t+d)^2$; (2) when $\La\to 0$, $d\to 0$, but 
$\La^5/d^2\to \mbox{constant}$, the breaking pattern is
 $USp(4) \to \widehat{U(1)}\times 
USp(2)$. 
The characteristic function in this case
becomes $P_2(t)=t(t+\La^5 m/d^2)$. 
Thus, we have a smooth transition from 
$\widehat{U(2)}\times USp(0)$ to $\widehat{U(1)}\times USp(2)$.

To count the number of vacua, using the relation
\bean
F_3(t)= W'(x)^2+{\cal O}(t)= t(t-m^2)^2+{\cal O}(t)
\Rightarrow -2m^2= a.
\nonu
\eean
There exist five  solutions.
\footnote{The mathematica calculation will give 
six solutions,
but one of them is that $d=0$ which is not the real solutions for our
problem because $d$ is not equal to zero at the beginning. 
Another easy way to count the number of vacua correctly is to use the numerical
method. } Three of them belong to $\widehat{U(2)}\times USp(0)$
by $(2N-N_f)=3$, and  two of them to $\widehat{U(1)}\times USp(2)$
where $U(1)$ is at the non-baryonic $r=0$ branch having 
one vacuum and $USp(2)$ 
contributes to
$(N+1)=2$.

To get $USp(4)\to USp(4)$, we require another double root or the curve
should be (\ref{example-4})
\bean
ty^2= t (t+a) (t^2 +b t+c)^2.
\nonu
\eean
There are five equations for five variables so 
all the parameters can be fixed. 
Two of them belong
to the breaking pattern $\widehat{U(1)}\times USp(2)$ 
where $\widehat{U(1)}$ is at the baryonic $r=0$ branch
and the $USp(2)$ gives two vacua $(N_0+1)=2$. 
Therefore, we have the same counting 
both in the strong and weak coupling analyses.  
The other three solutions belong to
the case $USp(4)\to USp(4)$, which matches  the counting of vacua in the 
weak coupling analysis $(N+1)=3$.

However, we still need to discuss the smooth transition carefully as
we did for $SO(5)$ with $N_f=1$ at the baryonic $r=0$ branch 
\cite{afo}. 
Carefully solving the problem, we find
\bean
s_1 & = & {-a-2b\over 2},~~~~~~s_2 = {-a(a-4 b)\over 8}+c, \\
b & = & a-{256 \La^{10}\over a^4},~~~~ c={3 a^5 -2048 \La^{10}
\over 16 a^3},\\
0 & = & m+{16 \La^5 \over a^2}-{a^3\over 64 \La^5}.
\nonu
\eean 
From these results we see the two classical limits: 
(1) when $\La^5/ a^2\to \mbox{constant}$, it gives
the breaking pattern $\widehat{U(1)}\times USp(2)$, and 
(2) as $a^3/\La^5\to 
\mbox{constant}$, it provides
$USp(4)$. Thus, both the case $USp(4)$ and 
the case $\widehat{U(1)}\times USp(2)$
where $\widehat{U(1)}$ is at the $r=0$ baryonic branch are smoothly connected.

\subsubsection{$N_f=2$ }

The curve is given by 
\bea
t y^2= \left[ t (t^2-s_1t +s_2)+ 2\La^4 m^2 \right]^2-4 
\La^{8}(t-m^2)^2.
\nonu
\eea

$\bullet$ {\bf Baryonic and Non-baryonic $r=1$ branches}

First let us consider the $r=1$ branch. There are two cases: one is
$\widehat{U(2)}$ at $r=1$ non-baryonic branch and the other, $\widehat{U(1)}$ 
at $r=N_f-N=1$ baryonic branch. For  $r=1$ non-baryonic branch,
 we require the factor
$(t-m^2)^2$ and the curve to be  factorized as follows (\ref{example-1}):
\bean
t y^2= t F_3(t) (t-m^2)^2 =t (t^3 +a t^2+ bt+c) (t-m^2)^2.
\nonu
\eean
The solution is given by
\bean
s_2 & = & -2 \La^4-m^4+m^2 s_1,~~~a=2(m^2-s_1), \nonu \\
b & = & -4 \La^4+m^4-2 m^2 s_1+s_1^2,
~~~c=-4(\La^4 m^2-\La^4 s_1).
\nonu
\eean
To fully determine the solution, we need to use $a=-2 m^2$ as we have observed
previously which
determines $s_1=2 m^2$. There is only one vacuum which matches the
counting of  the $r=1$ branch by $(2N-N_f)/2=1$
for $r=N_f/2$ \cite{bfhn,afo}. 
To get the baryonic $r=1$ branch, the curve should be factorized as,
from (\ref{example-3}),
\bean
ty^2= t(t+a) (t+b)^2 (t-m^2)^2.
\nonu
\eean
It is easy to see by 
mathematica computation that there are two solutions giving
a breaking pattern $USp(4) \to \widehat{U(1)}\times USp(2)$ with 
$\widehat{U(1)}$ at the
$r=1$ baryonic branch. The two vacua come from the $USp(2)$ factor,
which is also realized in the weak coupling analysis by $(N+1)=2$.

$\bullet$ { \bf Baryonic and Non-baryonic $r=0$ branches}

Next we will discuss the $r=0$ branch. In this case, except for the
$r=0$ non-baryonic branch for both $\widehat{U(2)}$ and 
$\widehat{U(1)}$, there is
also the $r=0$ baryonic branch for  $\widehat{U(2)}$. For the $r=0$ 
non-baryonic branch,
we require the factorization to be
\bean
t y^2= t F_3(t) H^2_1(t) = t (t^3 +a t^2+ bt+c) (t+d)^2.
\nonu
\eean
The first kind of solution 
is given by
\bean
s_1=-2d,~~~s_2=d^2-2\La^4,~~~a=2d,~~~b=d^2-4 \La^4,~~~c=4 \La^4 m^2.
\nonu
\eean
This gives $USp(4) \to \widehat{U(2)}$ at the classical limit. 
To count the number of vacua, using the 
relationship
\bean
 -2m^2= a
\nonu
\eean
we get one vacuum. Some remarks are in order.
Although the curve has a factor of  $(t+d)^2=(t-m^2)^2$ and  it seems to
belong to the $r=1$ branch, it is a  kind of an illusion and we should
take it to be the $r=0$ branch. Similar phenomena have been observed 
previously in
the $N_f=0$ case.

The second kind of solution is given by
\bean
s_1 & = &{-2 d - \frac{4 {\La}^4 m^2}{d^2}},~~~~
s_2 = {d^2 + 2 {\La}^4 + 
      \frac{8 {\La}^4\,m^2}{d}}, \\
a & = & {2 d + \frac{8 {\La}^4 m^2}{d^2}},~~~~b=
{d^2 + 4 {\La}^4 + 
      \frac{16 {\La}^4 m^2}{d} + 
      \frac{16 {\La}^8 m^4}{d^4}},\\
c & = & {\frac{4 {\La}^4 m^2\,
        \left( d^3 + 4 d {\La}^4 + 
          8 {\La}^4 m^2 \right) }{d^3}}.
\nonu
\eean
There are the following limits: 
(1) when $\La\to 0$, but $d\to \mbox{constant}$,
it produces $USp(4) \to USp(0) 
\times \widehat{U(2)}$;  (2) when $\La\to 0$, $d\to 0$ and
$\La^4/d^2\to \mbox{constant}$,  there exists 
$USp(4) \to \widehat{U(1)}\times 
USp(2)$. Thus, we have the smooth transition from 
$USp(0) \times \widehat{U(2)}$ to $\widehat{U(1)}\times USp(2)$.
To count the number of vacua, we use the fact $a=-2 m^2$ which gives 
three solutions for 
$d$. Two of  them give the symmetry breaking 
$USp(4) \to \widehat{U(1)}\times USp(2)$
where $\widehat{U(1)}$ is at the $r=0$ non-baryonic branch. One of them 
gives $USp(4) \to USp(0) \times \widehat{U(2)}$. 
Therefore, by combining with the one vacuum
from the first kind of solution, we get two vacua for $\widehat{U(2)}$
at the $r=0$ non-baryonic branch  which is exactly that in the 
weak coupling analysis obtained through the relation  $(2N-N_f)=2$.

Finally we require 
\bean
ty^2= t (t+a) (t+b)^2 (t+c)^2
\nonu
\eean
to get the  $USp(4)\to USp(4)$. Since it is at the $r=0$ branch, it is
equal to
\bean
& & t (t^2-s_1t +s_2)+ 2\La^4 m^2+2 \La^4 (t-m^2)= t (t+b)^2,\\
& & t (t^2-s_1t +s_2)+ 2\La^4 m^2-2 \La^4 (t-m^2)=(t+a)(t+c)^2.
\nonu
\eean 
Here the first relation can be understood 
that the left hand side does not have a $t$-independent term, 
it does have a factor
$t$ and therefore
it is a consistent equation.
The solution is given by
\bean
s_1 & = & -2 c-{4\La^4 m^2 \over c^2},~~~~s_2= -2 \La^4+
{ (c^3+2 \La^4 m^2)^2 \over c^4},\\
a & = & {4 \La^4 m^2 \over c^2},~~~~~b= c+ {2 \La^4 m^2\over c^2},\\
0 & =& c^4+ c^3 m^2-\La^4 m^4.
\nonu
\eean
There are four solutions. Keeping the first and second terms of the
last equation, we get one solution $c,b\sim -m^2$ which gives
a breaking pattern $USp(4) \to USp(0) \times 
\widehat{U(2)}$ where $\widehat{U(2)}$
is at the $r=0$ baryonic branch. The vacuum number one 
is also realized in the
weak coupling analysis \cite{bfhn,afo}. Keeping the second and third terms
of the last equation, we get three solutions $c\sim \La^{4/3}$ 
which give a pattern 
$USp(4)\to USp(4)$. Finally, there is a smooth transition from
$USp(4) \to USp(0) \times \widehat{U(2)}$  (where $\widehat{U(2)}$
is at the $r=0$ baryonic branch) to $USp(4)\to USp(4)$,
which matches  the counting of vacua in the 
weak coupling analysis $(N+1)=3$.

\subsubsection{$N_f=3$}

The curve can be written as 
\bea
t y^2= \left[ t (t^2-s_1t +s_2)+ 2\La^3 m^3 
\right]^2+4 \La^{6}(t-m^2)^3.
\nonu
\eea

$\bullet$ {\bf Baryonic and Non-baryonic $r=1$ branches}

First let us consider the $r=1$ branch. There will be an $r=1$ non-baryonic
branch for $\widehat{U(2)}$ and an $r=1$ baryonic branch for both 
$\widehat{U(1)}$ and 
$\widehat{U(2)}$.  For the $r=1$ non-baryonic branch, 
the curve should be, from (\ref{example-1}),
\bean
t y^2= t F_3(t) (t-m^2)^2 =t (t^3 +a t^2+ bt+c) (t-m^2)^2.
\nonu
\eean
The solution is given by
\bean
s_2 & = & -m ( 2\La^3+M^3-m s_1),~~~~~c= 4 \La^3( \La^3-m^3+ m s_1),\\
a & = & 2 (m^2-s_1),~~~~b=-4 \La^3 m + (m^2-s_1)^2.
\nonu
\eean
Using the relation $a=-2 m^2$, we find that $s_1=2 m^2$ and the curve
gives one vacuum for $USp(4) \to USp(0) \times 
\widehat{U(2)}$ at the $r=1$ branch,
which is also observed in the weak coupling analysis by counting $(2N-N_f)=1$. 

To get the $r=1$ baryonic branch for 
$\widehat{U(1)}\times USp(2)$ we need to have the following 
factorized curve (\ref{example-3})
\bean
ty^2= t(t+a) (t+b)^2 (t-m^2)^2.
\nonu
\eean
There are three solutions given by 
\bean
a^3- 64 \La^6+ 16 a \La^3 m=0,~~~~b={ a^3+2 a^2 m^2-32 
\La^3( \La^3+m^3)
\over 2 a(a+4m^2)}.
\nonu
\eean
Two of them give a breaking pattern $USp(4) \to \widehat{U(1)}\times USp(2)$ 
at the $r=1$ baryonic 
branch (the weak coupling analysis provides the number of vacua as two from the
$USp(2)$ factor by $(N+1)=2$) and the remainding one, 
$USp(4) \to USp(0) \times \widehat{U(2)}$ where 
$\widehat{U(2)}$ is at the $r=1$ baryonic branch, which is coincident with the 
number of vacua in the weak coupling approach. Thus,
we see the smooth transition of the $r=1$ baryonic branch
between $USp(0) \times \widehat{U(2)}$ and $\widehat{U(1)}\times USp(2)$.

$\bullet$ {\bf Non-baryonic $r=0$ branch}

Now let us discuss the $r=0$ branch. There is no  baryonic 
$r=0$ branch. The curve is given by
\bean
t y^2= t F_3(t) H^2_1(t) = t (t^3 +a t^2+ bt+c) (t+d)^2.
\nonu
\eean
The solutions are
\bean
s_1 & = & \frac{\pm {\La}^3 \left( d - 2 m^2 \right) 
     {\sqrt{d + m^2}} - 2
     \left( d^3 + {\La}^3 m^3 \right) }{d^2},\\
s_2 & = & \frac{d^3 + 4 {\La}^3 m^3 \pm
    {\La}^3 {\sqrt{d + m^2}}
     \left( d + 4 m^2 \right) }{d}.
\nonu
\eean
There are the following limits: (1) when $\La\to 0$, but $d\to 
\mbox{constant}$,
there exists a breaking pattern 
$USp(4) \to USp(0) \times 
\widehat{U(2)}$; (2) when $\La\to 0$, $d\to 0$, but 
$\La^3/d^2\to \mbox{constant}$, so that $USp(4) \to \widehat{U(1)}\times 
USp(2)$. Thus, we have a smooth transition from 
$USp(0) \times  
\widehat{U(2)}$ to $\widehat{U(1)}\times USp(2)$. To count the
number of vacua, we set $a=-2m^2$ as before and solve $s_1,s_2, b,c$ in terms of
 $d$
with the following constraint (we have neglected the case where $d=-m^2$)
\bean
0 = d^4 - d {\La}^6 + 2 d^3 m^2 + 
  3 {\La}^6 m^2 + 
  4 d {\La}^3 m^3 + d^2 m^4 + 
  4 {\La}^3 m^5.
\nonu
\eean
There are four solutions. Two  of them give  
$USp(4) \to \widehat{U(1)}\times 
USp(2)$ (in the weak coupling approach the number two comes from the 
$USp(2)$ part while the number of vacua for the $\widehat{U(1)}$ part gives one
vacuum from the relation $(N-r)=1$) and another two, 
$USp(4) \to USp(0) \times \widehat{U(2)}$. 
It is noteworthy that for the $r=0$ non-baryonic branch of
$ \widehat{U(2)}$ with $N_f=3$, the counting is given by 
$(N-r)=2$ from \cite{bfhn,afo} in the weak coupling approach
instead of in the $(2N-N_f)=1$.

Finally we require the following factorization curve
\bean
ty^2= t (t+a) (t^2 +b t+c)^2.
\nonu
\eean
It is hard to solve analytically, but numerically there exist
twelve solutions. However, six of them have $s_1\to m^2$, three of
them have $s_1\to 2 m^2$, and  three of them have $s_1\to 0$,
which is the one we are looking for and give $USp(4)\to USp(4)$. 
 The counting of vacua in the 
weak coupling analysis gives $(N+1)=3$, which is the same as the one in the 
above strong  coupling analysis.

\subsubsection{$N_f=4$}

The curve is
\bea
t y^2= \left[ t (t^2-s_1t +s_2)+ 2\La^2 m^4 \right]^2-4 
\La^{4}(t-m^2)^4.
\nonu
\eea

$\bullet$ {\bf Baryonic $r=2$ branch}

There is only one baryonic $r=2$ branch for $\widehat{U(2)}$. 
The curve is given by (\ref{example-3})
\bean
t y^2= t (t+a) (t-m^2)^4.
\nonu
\eean
There is one solution:
\bean
a= 4 \La^2,~~~~s_1=2m^2-2 \La^2,~~~~s_2= m^4- 4 \La^2 m^2
\nonu
\eean
which gives a breaking pattern $USp(4) \to USp(0) \times \widehat{U(2)}$ 
at the  baryonic $r=2$ branch. In the weak coupling analysis \cite{bfhn,afo}
the number of vacua is
also one because it is a baryonic branch.

$\bullet$ {\bf Non-baryonic $r=1$ branch}

For the $r=1$ branch, it is non-baryonic and the curve should 
be (\ref{example-1})
\bean
t y^2= t (t^3 +a t^2+ bt+c) (t-m^2)^2.
\nonu
\eean
The solution is given by:
\bean
s_2 & = & -m^2( 2\La^2 + m^2-s_1),~~~~~c= 4\La^2 m^2
( 2\La^2 -m^2+s_1),\\
a & = & 2(m^2-s_1),~~~~~b=-4 \La^4 -4 \La^2 m^2+(m^2-s_1)^2.
\nonu
\eean
Using $a=-2m^2$, we see that there is one vacuum for 
$USp(4) \to USp(0) \times \widehat{U(2)}$, 
which can be checked also by the 
counting $(N-r)=(2-1)=1$ in the weak coupling analysis. We did not
see the breaking pattern 
$USp(4) \to \widehat{U(1)}\times USp(2)$ because for the
$\widehat{U(1)}$ factor, we have $(N-r)=0$.

$\bullet$ {\bf Non-baryonic $r=0$ branch}

For the $r=0$ branch, it is also non-baryonic and the curve should be
\bean
t y^2= t F_3(t) H^2_1(t) = t (t^3 +a t^2+ bt+c) (t+d)^2.
\nonu
\eean
There are two kinds of solutions. The first kind of solution 
is given by
\bean
s_1 & = & -2(d+\La^2),~~~~s_2= d^2-4 \La^2 m^2,~~~~c= 4\La^2 m^4,\\
a & = & 2(d+2 \La^2),~~~~b=d^2-8 \La^2 m^2.
\nonu
\eean
It gives one vacuum for the breaking  $USp(4) \to USp(0) 
\times \widehat{U(2)}$. 
The second
kind of solution is given by
\bean
s_1 & = & -2d+2\La^2 -{4 \La^2 m^4 \over d^2},~~~~
s_2 = d^2 + 4 \La^2 m^2 +{8 \La^2 m^4 \over d},\\
a & = & 2(d-2\La^2+{4\La^2 m^4 \over d^2}),~~~~b=
d^2+ 8\La^2 m^2 +{16 \La^2(d-\La^2) m^4\over d^2}+{16 \La^4 m^8 \over
d^4},\\
c & = & { 4\La^2 m^4 (d^3+8 d \La^2 m^2+ 8 
\La^2 m^4)\over d^3}.
\nonu
\eean
There exist the  following limits: (1) when
 $\La\to 0$, but $d\to \mbox{constant}$,
there is a breaking pattern 
$USp(4) \to USp(0) \times 
\widehat{U(2)}$; (2) when $\La\to 0$, $d\to 0$, but 
$\La^2/d^2\to \mbox{constant}$, we have $USp(4) \to \widehat{U(1)}\times 
USp(2)$. Thus, we have the smooth transition from 
$USp(0) \times 
\widehat{U(2)}$ to $\widehat{U(1)}\times USp(2)$.  To count the
number of vacua, we set $a=-2m^2$ as we did before and find three solutions. 
Two of them give a breaking pattern 
$\widehat{U(1)}\times USp(2)$ where the $\widehat{U(1)}$
factor is with the counting $(N-r)=1$ and $USp(2)$ has two vacua. 
The remainding one gives
$USp(0) \times 
\widehat{U(2)}$ and by combining the one vacuum from the first kind
of solutions we get a total of 
two vacua, which is also computed by the 
counting $(N-r)=2$ in the weak coupling analysis.

Finally we require the following factorization
\bean
& & t (t^2-s_1t +s_2)+ 2\La^2 m^4-2 \La^2 (t-m^2)^2= t (t+b)^2,\\
& & t (t^2-s_1t +s_2)+ 2\La^2 m^4+2 \La^2 (t-m^2)^2=(t+a)(t+c)^2.
\nonu
\eean 
The solution is given by some function of $c$ 
while the parameter $c$ satisfies the
following constraint
\bean
0 = -c^3 + c^2 \La^2-2 c \La^2 m^2+\La^2 m^4.
\nonu
\eean
Obviously there are three solutions which provide the three vacua of
$USp(4)\to USp(4)$.

\subsubsection{$N_f=5$}

The curve can be described by:
\bea
t y^2= \left[ t (t^2-s_1t +s_2)+ 2\La m^5 \right]^2+4 \La^{2}
(t-m^2)^5.
\nonu
\eea

$\bullet$ {\bf Baryonic $r=2$ branch}

The $r=2$ branch is baryonic and the curve is from (\ref{example-3})
\bean
t y^2= t (t+a) (t-m^2)^4.
\nonu
\eean
The solution is given by:
\bean
s_1= 2m(-\La+m),~~~~s_2= m^3(-4 \La+m),~~~~a=4 \La(\La+m)
\nonu
\eean
which gives the one vacuum of 
$USp(4) \to USp(0) \times 
\widehat{U(2)}$ at the  baryonic $r=2$ branch, which is also
consistent with  the count in the weak coupling analysis.

$\bullet$ {\bf Non-baryonic $r=1$ branch}

For the $r=1$ branch, it is non-baryonic and the curve should be
(\ref{example-1})
\bean
t y^2= t (t^3 +a t^2+ bt+c) (t-m^2)^2.
\nonu
\eean
The solution is given by
\bean
s_2 & = & m^2(-m(2\La+m)+s_1),~~~~~a= 2 (2\La^2+m^2-s_1),\\
b& = & -12 \La^2 m^2- 4 \La m^3+(m^2-s_1)^2,~~~~c= 
4\La m^3( 3\La m-m^2+
s_1).
\nonu
\eean
Using $a= -2 m^2+4 \La^2$, we get one vacuum for 
$USp(4) \to USp(0) \times \widehat{U(2)}$ that can be interpreted in the 
weak coupling analysis by counting $(N-r)=1$. We did not
see the $USp(4) \to \widehat{U(1)}\times USp(2)$ because for the
$\widehat{U(1)}$ factor we have no vacua $(N-r)=0$.

$\bullet$ {\bf Non-baryonic $r=0$ branch}

For $r=0$ branch, it is also non-baryonic and the curve should be
\bean
t y^2= t F_3(t) H^2_1(t) = t (t^3 +a t^2+ bt+c) (t+d)^2.
\nonu
\eean
The solutions are given by
\bean
s_1 & = & \frac{\pm \La \left( 3 d - 2 m^2 \right) 
     {\left( d + m^2 \right) }^{\frac{3}{2}} - 
    2 \left( d^3 + \La m^5 \right) }{d^2},\\
s_2 & = & \frac{d^3 + 4 \La m^5 \mp  
    \La \left( d - 4 m^2 \right) 
     {\left( d + m^2 \right) }^{\frac{3}{2}}}{d}.
\nonu
\eean
There are the following limits: (1) when $\La\to 0$, but $d\to 
\mbox{constant}$,
we see $USp(4) \to USp(0) \times 
\widehat{U(2)}$; (2) as $\La\to 0$, $d\to 0$, but 
$\La/d^2\to \mbox{constant}$, so that the following 
breaking pattern appears $USp(4) \to \widehat{U(1)}\times 
USp(2)$. Thus, we have the smooth transition from 
$USp(0) \times \widehat{U(2)}$ to $\widehat{U(1)}\times USp(2)$. To 
count the number of vacua,
setting $a=-2m^2+4 \La^2$, we solve other parameters  
in terms of $d$ with the
constraint
\bean
0 = d^4 + d\La m^4
   \left( 5\La + 4m \right)  + 
  \La m^6 \left( 15 \La + 4 m
     \right)  + d^3 \left( -9 {\La}^2 + 
     2 m^2 \right)  + d^2 
   \left( -15 {\La}^2 m^2 + m^4 \right).
\nonu
\eean
Among these four solutions, two of them give a breaking pattern 
$USp(4) \to USp(0) \times
\widehat{U(2)}$
which can be realized in the weak coupling analysis by 
counting $(N-r)=2$, and in the two others, 
$USp(4) \to \widehat{U(1)}\times 
USp(2)$ where two vacua  are 
also obtained from the analysis
of weak coupling $(N_1-r) \times (N_0+1)=2$.

Finally we set 
\bean
ty^2= t (t+a) (t^2 +b t+c)^2.
\nonu
\eean
It should give three vacua, but we are unable to show 
it  numerically. 

We summarize the results in Table 1 by writing the flavors $N_f$,
symmetry breaking patterns, various branches, the power of $t$ in the
curve, $U(1)$ at the IR, the number of vacua, and the possible smooth
connection.

\begin{table} 
\begin{center}
\begin{tabular}{|c|c|c|c|c|c|c|} \hline
 $N_f$  &  Group  & Branch &  Power of $t(=x^2)$ &  $U(1)$ & 
Number of vacua & Connection
 \\ \hline
 0 & ${USp(4)}$  & $(C)$ &  $t^1$ & 0 & 3 & \\ \cline{2-7}
 & $USp(2) \times {U(1)}$  & $(C,0_{NB})$ &  $t^1$ & 1 & 2 & A \\ \cline{2-7}
& $USp(0) \times {U(2)}$ & $(C,0_{NB})$ & $t^1$ 
& 1 & 1 & A \\ \cline{3-7}
&  & $(C,0_{NB})$ & $t^1$ & 1 & 1 &  \\ \hline 
1 & ${USp(4)}$  & $(C)$ &  $t^1$ & 0 & 3 & B \\ \cline{2-7}
 & ${USp(2)}\times \widehat{U(1)}$  & $(C,0_{B})$ &  $t^1$ & 0 & 2 &B \\ \cline{3-7}
 &   & $(C,0_{NB})$ &  $t^1$ & 1 & 2 & C \\ \cline{2-7}
& $USp(0) \times \widehat{U(2)}$ & $(C,0_{NB})$ & $t^1$ & 1 & 3 & C \\ \hline 
2 & ${USp(4)}$  & $(C)$ &  $t^1$ & 0 & 3 &D \\\cline{2-7}
 & ${USp(2)}\times \widehat{U(1)}$  & $(C,1_{B})$ &  $t^1$ & 0 & 2 & \\\cline{3-7}
 &   & $(C,0_{NB})$ &  $t^1$ & 1 & 2 &E \\\cline{2-7}
 & $USp(0) \times \widehat{U(2)}$ & $(C,0_{NB})$ & $t^1$ & 1 & 1 &E \\ \cline{3-7}
 &  & $(C,1_{NB})$ & $t^1$ & 1 & 1 & \\ \cline{3-7}
 &  & $(C,0_{NB})$ & $t^1$ & 1 & 1 & \\ \cline{3-7}
& & $(C,0_{B})$ & $t^1$ & 0 & 1 &D \\ \hline 
3 & ${USp(4)}$  & $(C)$ &  $t^1$ & 0 & 3 &  \\\cline{2-7}
 & ${USp(2)}\times \widehat{U(1)}$  & $(C,1_{B})$ &  $t^1$ & 0 & 2 & F \\\cline{3-7}
 &   & $(C,0_{NB})$ &  $t^1$ & 1 & 2 & G \\\cline{2-7}
 & $USp(0) \times \widehat{U(2)}$ & $(C,1_{NB})$ & $t^1$ & 1 & 1 &  \\ \cline{3-7}
 && $(C,1_{B})$ & $t^1$ & 0 & 1 & F \\ \cline{3-7}
&  & $(C,0_{NB})$ & $t^1$ & 1 & 2 & G \\ \hline    
4 & ${USp(4)}$  & $(C)$ &  $t^1$ & 0 & 3 & \\ \cline{2-7}
 & ${USp(2)}\times \widehat{U(1)}$  & $(C,0_{NB})$ &  $t^1$ & 1 & 2 &H \\ \cline{2-7}
 & $USp(0) \times \widehat{U(2)}$ & $(C,2_{B})$ & $t^1$ & 0 & 1 & \\ \cline{3-7}
  &  & $(C,1_{NB})$ & $t^1$ & 1 & 1 & \\ \cline{3-7}
   &  & $(C,0_{NB})$ & $t^1$ & 1 & 1 & \\ \cline{3-7}
&  & $(C,0_{NB})$ & $t^1$ & 1 & 1 &H \\ \hline 
5 & ${USp(4)}$  & $(C)$ &  $t^1$ & 0 & $3^{\ast}$ & \\\cline{2-7}
 & ${USp(2)}\times \widehat{U(1)}$  & $(C,0_{NB})$ &  $t^1$ & 1 & 2 &I \\\cline{2-7}
&  $USp(0) \times \widehat{U(2)}$  & $(C,2_{B})$ &  $t^1$ & 0 & 1 & \\\cline{3-7}
&    & $(C,1_{NB})$ &  $t^1$ & 1 & 1 & \\\cline{3-7}
& & $(C,0_{NB})$ & $t^1$ & 1 & 2 & I \\ \hline 
\end{tabular} 
\end{center}
\caption{\sl The summary of the phase 
structure of the $USp(4)$ gauge group with massive flavors. 
The flavors are charged under the $U(N_i)$ group.
Here we use $C$ for the Chebyshev  branch for the $USp(2N_i)$ factor
and $r_{NB}/r_{B}$ for the $r$-th non-baryonic or baryonic  branch. 
In this table, we list the power of $t$ that is $t^1$ 
and the $U(1)$ which is
present in the nonbaryonic branch, at the
IR. 
They are  indices to see whether the 
two phases could have smooth
transition. 
Note  the explicit presence of the $USp(0)$ group factor in the 
second column. The same capital letters in the last column denote
 the smooth connected phases between them. We could not find  three 
vacua denoted by $3^{\ast}$ above.}
\label{massivetableusp4}
\end{table}

\subsection{ $USp(6)$ case}

\indent

The calculation becomes more complicated as we increase the rank of the 
gauge group for $USp(2N_c)$. For the $USp(6)$ gauge group,
let us deal with only the even number of flavors because  there
are no new and interesting phenomena for odd number of flavors.

\subsubsection{$N_f=0$}

The curve is 
\bean
 ty^2=\left[ t (t^3 -s_1 t^2+ s_2 t-s_3)+ 2\La^8 \right]^2- 4 
\La^{16}.
\nonu
\eean

$\bullet$ {\bf Non-baryonic $r=0$ branch}

First we require that 
\bean
t y^2 & = & t F_3(t) H^2_2(t). 
\nonu
\eean
There are two kinds of solutions. The first kind of solution
is given by, from the simple further factorization, 
\bean
t (t^3 -s_1 t^2+ s_2 t-s_3)+ 2\La^8 -2 \La^8 & = &  t F_3(t), \\
t (t^3 -s_1 t^2+ s_2 t-s_3)+ 2\La^8 +2 \La^8 & = &  H_2(t)^2=
(t^2+a t +b)^2.
\nonu
\eean
The solution is given by:
\bean
s_1 =-2 a,~~~s_2= a^2\mp 4\La^4,~~s_3= \pm 4 
a\La^4,~~~~b=\mp 2\La^4
\nonu
\eean
so that the function $F_3(t)$ can be written as
\bean
F_3(t)= (t+a)( t(t+a)^2 \mp \La^4).
\nonu
\eean
Using the relationship of $F_3(t)$ and $W'(x)$, we get $a=-m^2$. 
There 
are two vacua for $USp(6) \to U(2) \times 
USp(2)$. The second kind of solution is given by: 
\bean
t (t^3 -s_1 t^2+ s_2 t-s_3)+ 2\La^8 -2 \La^8 & = &  t(t+a) (t+b)^2, \\
t (t^3 -s_1 t^2+ s_2 t-s_3)+ 2\La^8 +2 \La^8 & = & 
(t^2+c t+d)(t+e)^2.
\nonu
\eean
The solution is given by
\bean 
s_1 & = & \frac{-3 \left( b + e \right) }{2} - 
  \frac{2 {\La}^8}{\left( b - e \right) e^2},~~~
s_2=b \left( 3 e + \frac{4 {\La}^8}
     {\left( b - e \right) e^2} \right),~~~
s_3 =\frac{b^2 \left( b - 3 e + 
      \frac{4 {\La}^8}
       {e^2 \left( -b + e \right) } \right) }{2},\\
a & = & \frac{-b}{2} + \frac{3 e}{2} + 
  \frac{2 {\La}^8}{\left( b - e \right) e^2},~~~
c=\frac{3 b}{2} - \frac{e}{2} + 
  \frac{2 {\La}^8}{\left( b - e \right) e^2},~~~
d=\frac{4 {\La}^8}{e^2},\\
0 & = & \frac{{\left( -b + e \right) }^3}{2} + 
  \frac{2 \left( b - 3 e \right) {\La}^8}
   {e^2}.
\nonu
\eean
There are the following limits: (1) as $\La\to 0$, $b\sim e\neq 0$, so
there is a relation $(b-e)\sim \La^{8/3}$.
It gives $USp(6) \to USp(0) \times U(3)$;
(2)  as $\La\to 0$, $b\neq 0$, we have $\La^8/e^2 \sim b^2/4$. It gives
$USp(6) \to U(2) \times 
USp(2)$; (3) when $\La\to 0$, $b,e\to 0$, but $\La^8/(e^2(b-e))\neq 0$,
it gives $USp(6) \to U(1) \times 
USp(4)$. Thus, we see the smooth transition among all of these three
phases.

To count the number of vacua, we find $F_3(t)$ and get 
a relation
\bean
b + e + \frac{4 {\La}^8}
   {\left( b - e \right) e^2}=-2 m^2.
\eean
Combining with this constraint, we get the following equation for $b$
\bean
0=-4 {\La}^{16} + 
  {\left( b + m^2 \right) }^2
   \left( b^6 + 28 b^2 {\La}^8 + 
     3 b \left( b^4 + 18 {\La}^8 \right) 
      m^2 + 3 \left( b^4 + 9 {\La}^8 \right) 
      m^4 + b^3 m^6 \right).
\nonu
\eean
There are eight solutions. Three of them give 
$USp(6) \to USp(0) \times U(3)$, which is consistent with the counting of
weak coupling analysis. Another three of them give
$USp(6) \to U(1) \times 
USp(4)$ where the number of vacua three comes from the $USp(4)$ part
in the weak coupling approach. The remainding two give $USp(6) 
\to U(2) \times 
USp(2)$. Combining the two vacua from the first kind of
factorization we have found before 
with this gives four vacua of this phase  (in the weak coupling 
analysis two comes from $USp(2)$ and two does from 
$U(2)$).

Finally we consider the case which is relevant to the 
pattern $USp(6) \to USp(6)$
\bean
t (t^3 -s_1 t^2+ s_2 t-s_3)+ 2\La^8 -2 \La^8 & = &  t(t+a) (t+b)^2, \\
t (t^3 -s_1 t^2+ s_2 t-s_3)+ 2\La^8 +2 \La^8 & = & (t+c)^2(t+d)^2.
\nonu
\eean
Solving the factorization will give us eight solutions, but putting
back to  curve we find only four different curves. Thus, the
four vacua match the prediction by $(N+1)=4$ for $USp(6)$ factor 
from the analysis of
weak coupling.

\subsubsection{$N_f=2$}

The curve is given by
\bean
 ty^2=\left[ t (t^3 -s_1 t^2+ s_2 t-s_3)+ 2\La^6 m^2 \right]^2- 
4 \La^{12} (t-m^2)^2.
\nonu
\eean

$\bullet$ {\bf Baryonic and Non-baryonic $r=1$ branches}

The $r=1$ branch is as follows: $\widehat{U(3)}$ and 
$\widehat{U(2)}$ have
 only non-baryonic branch while $\widehat{U(1)}$ is at 
the baryonic branch.
To do this, first we require 
\bean
 t (t^3 -s_1 t^2+ s_2 t-s_3)+ 2\La^6 m^2= (t-m^2) 
(t^3 -u_1 t^2+ u_2 t-u_3).
\nonu
\eean
This can be solved as
\bean
s_1= m^2+u_1,~~~s_2= m^2 u_1+u_2,~~~~s_3= 2\La^6+m^2 u_2,~~~~
u_3= 2\La^6.
\nonu
\eean
Now we can have the following factorization
\bean
  (t^3 -u_1 t^2+ u_2 t -2\La^6)^2-4 \La^{12}
 & = &  \left[ t(t^2- u_1 t+ u_2) -2 \La^6\right]^2-4 \La^{12}
\nonu \\
& = & t (t^3+ a t^2+b t+c) (t+d)^2
\nonu
\eean
which remarkably is the $r=0$ nonbaryonic branch 
of $USp(4)$ without flavors 
(\ref{flavorless4}).
From these results, we conclude immediately that: (1) 
there is smooth transition between $USp(6) \to USp(0) 
\times \widehat{U(3)}$
and $USp(6) \to \widehat{U(2)}\times  USp(2)$; (2) there are 
two vacua for $USp(6) \to USp(0) \times 
\widehat{U(3)}$, which match the counting
$(2N_c-N_f)/2=2$ in the weak coupling analysis; and (3)  there are 
two vacua for $USp(6) \to \widehat{U(2)}\times  USp(2)$, which 
match the counting $2\times (2N-N_f)/2=2$; (4)
If we require the extra double root, we get three vacua for
$USp(6) \to \widehat{U(1)}\times  USp(4)$ where $\widehat{U(1)}$ 
is at the $r=1$ baryonic branch and the number three in the weak 
coupling analysis comes from the $USp(4)$ factor.

$\bullet$ {\bf Baryonic and Non-baryonic $r=0$ branches}

Now we will discuss the $r=0$ branch. There are non-baryonic branches for
$\widehat{U(3)}$, $\widehat{U(2)}$ and $\widehat{U(1)}$. 
For $\widehat{U(2)}$, there is also the baryonic
branch. First let us discuss the non-baryonic branch. For the
factorization $(2,0)$ where the numbers $(X,Y)$ are 
indicating the numbers of
double roots, it is given by
\bean
t (t^3 -s_1 t^2+ s_2 t-s_3)+ 2\La^6 m^2 + 2 \La^6 (t-m^2) & = & 
t F_3(t), \\
t (t^3 -s_1 t^2+ s_2 t-s_3)+ 2\La^6 m^2 - 2 \La^6 (t-m^2) & = & 
(t^2+a t+b)^2.
\nonu
\eean
The solutions are given by
\bean
s_1=-2a,~~~s_2=a^2\mp 4\La^3 m,~~~s_3= -2 \La^6\pm4 \La^3 a m,~~~
b=\mp 2\La^3 m
\nonu
\eean
which give the breaking $USp(6) \to \widehat{U(2)}\times  USp(2)$.
Calculating the $F_3(t)$, we get a relation 
$a=-m^2$ and there are two vacua.

For 
factorization $(1,1)$ the curve is  given by
\bean
t (t^3 -s_1 t^2+ s_2 t-s_3)+ 2\La^6 m^2 + 2 \La^6 (t-m^2) & = & 
t (t+a) (t+b)^2, \\
t (t^3 -s_1 t^2+ s_2 t-s_3)+ 2\La^6 m^2 - 2 \La^6 (t-m^2) & = & 
(t^2+c t+d)(t+e)^2.
\nonu
\eean
The solution is given by 
\bean
s_1 & = & \frac{-3 \left( b + e \right) }{2} - 
  \frac{2 {\La}^6 m^2}
   {\left( b - e \right) e^2},~~~
s_2=b \left( 3 e + \frac{4 {\La}^6 m^2}
     {\left( b - e \right) e^2} \right),~~~
d=\frac{4 {\La}^6 m^2}{e^2},\\
s_3 & = & \frac{4 {\La}^6 + 
    b^2 \left( b - 3 e + 
       \frac{4 {\La}^6 m^2}
        {e^2 \left( -b + e \right) } \right) }{2},~~~
a=\frac{-b}{2} + \frac{3 e}{2} + 
  \frac{2 {\La}^6 m^2}
   {\left( b - e \right) e^2}, \\
c & = & \frac{3 b}{2} - \frac{e}{2} + 
  \frac{2 {\La}^6 m^2}
   {\left( b - e \right) e^2}, ~~~
0  =  {(-b+e)^3-8 \La^6 \over 2}+2 {(b-3e) \La^6 m^2 
\over e^2}.
\nonu
\eean
There are the following limits: (1) when $\La\to 0$, $b\sim e\neq 0$, 
so it gives
$(b-e)\sim \La^{8/3}$ which implies  a breaking pattern 
$USp(6) \to USp(0) \times \widehat{U(3)}$;
(2) when $\La\to 0$, $b\neq 0$, we have a relation 
$\La^6 m^2/e^2 \sim b^2/4$. This gives
a breaking pattern $USp(6) \to \widehat{U(2)}\times USp(2)$; 
(3) when $\La\to 0$, $b,e\to 0$, but $\La^6 m^2/(e^2(b-e))\neq 0$,
it gives a symmetry breaking pattern $USp(6) \to \widehat{U(1)}\times 
USp(4)$. Thus we see the smooth transition among all of these 
three phases.

To count the number of vacua, 
we find the $F_3(t)$ and get the following 
relationship
\bean
-2 m^2= b + e + \frac{4 {\La}^6 m^2}
   {\left( b - e \right) e^2}.
\nonu
\eean
Combined with the above constraint  and eliminating $e$, we get 
an equation for $b$ of degree nine. 
Among these nine solutions, four give
 a breaking pattern $USp(6) \to USp(0) \times 
\widehat{U(3)}$ which is the same as 
the number $(2N-N_f)=4$; three give  
$USp(6) \to 
\widehat{U(1)}\times 
USp(4)$, which is the same as the counting $(N_1-r) \times (N_0+1)= 
1 \times 3 =3$; 
and the remainder two, plus the other two from the
previous $(2,0)$ factorization, 
for $USp(6) \to \widehat{U(2)}\times USp(2)$ which is also obtained 
from $(2N_1-N_f) \times (N_0+1)=2 \times 2=4$ in the weak 
coupling analysis.

Finally we discuss the following factorization
\bean
t (t^3 -s_1 t^2+ s_2 t-s_3)+ 2\La^6 m^2 + 2 \La^6 (t-m^2) & = & 
t (t+a) (t+b)^2, \\
t (t^3 -s_1 t^2+ s_2 t-s_3)+ 2\La^6 m^2 - 2 \La^6 (t-m^2) & = & 
(t+c)^2(t+d)^2.
\nonu
\eean
The solution is given by
\bean
s_1 & = & -a-2b,~~~s_2=b(2a+b),~~~~s_3=-a b^2+2 \La^6, \\
u_1 & \equiv & c+d={a\over 2}+b,~~~~u_2\equiv c d =\mp 2\La^3 m, \\
b & = & {a\over 2}-{16 \La^6 \over a^2},~~~~~
{a^2 \over \La^3}-{64 \La^3\over a}\mp 16 m=0.
\nonu
\eean
There are six solutions. Four of them have ${a^2 \over \La^3}\neq 0$
and give the four vacua of $USp(6) \to USp(6)$ which is the same as
the counting $(N+1)=4$. The remainder two 
have $ {64 \La^3\over a}\neq 0$ and give 
the breaking pattern 
$USp(6) \to \widehat{U(2)}\times  USp(2)$ where $\widehat{U(2)}$ is
at the $r=0$ baryonic branch (in the weak coupling analysis the number 
two comes from the factor $USp(2)$ with $(N+1)=2$). 
These two phases are smoothly connected.

\subsubsection{$N_f=4$}

The curve is 
\bean
 ty^2=\left[ t (t^3 -s_1 t^2+ s_2 t-s_3)+ 2\La^4 m^4 \right]^2- 
4 \La^{8} (t-m^2)^4.
\nonu
\eean

$\bullet$ {\bf Non-baryonic $r=0$ branch}

Now we discuss the $r=0$ branch. There are only non-baryonic 
branches for
$\widehat{U(3)}$, $\widehat{U(2)}$, and $\widehat{U(1)}$. 
For the factorization $(2,0)$ it is given by
\bean
t (t^3 -s_1 t^2+ s_2 t-s_3)+ 2\La^4 m^4 - 2 \La^4 (t-m^2)^2 & = & 
t F_3(t), \\
t (t^3 -s_1 t^2+ s_2 t-s_3)+ 2\La^4 m^4 + 2 \La^4 (t-m^2)^2 & = & 
(t^2+a t+b)^2.
\nonu
\eean
The solutions are given by
\bean
s_1=-2a,~~~s_2=a^2-2 \La^4-4 \eta \La^2 m^2 ,~~~s_3= 4\eta 
\La^2 m^2 (a-\eta \La^2),~~~
b=-2\eta \La^2 m^2
\nonu
\eean
which give the breaking $USp(6) \to \widehat{U(2)}\times  USp(2)$.
Calculating the $F_3(t)$, we get a relation 
$a=-m^2$ with two vacua.

For the
factorization $(1,1)$ type, the curve is  given by
\bean
t P_3(t)+ 2\La^4 m^4 - 2 \La^4 (t-m^2)^2 & = & 
t (t+a) (t+b)^2, \\
t P_3(t)+ 2\La^4 m^4 + 2 \La^4 (t-m^2)^2 & = & 
(t^2+c t+d)(t+e)^2.
\nonu
\eean
The solution is given by 
\bean
d&=&\frac{4\La^4 m^4}{e^2},\ \ \ 
a=\frac{e(b+e)}{2b}+\frac{\La^4 (e+m^2)(e+3m^2)}{be(b-e)}, \\
c&=&\frac{b(b+e)}{2e}+\frac{2\La^4(e+m^2)
(be-bm^2+4em^2)}{e^3(b-e)}, \\
0 & = & \frac{(b-e)^3}{2}+\frac{2\La^4}{e^2}(e+m^2)
(be+e^2-bm^2+3em^2).
\nonu
\eean
There are the following limits: (1) when $\La\to 0$, $b\sim e\neq 0$, 
so it gives
$(b-e)\sim \La^{4/3}$ which implies a symmetry breaking 
pattern $USp(6) \to USp(0) \times \widehat{U(3)}$;
(2)  when $\La\to 0$, $b\neq 0$, we have 
the following relation $\La^4 /e^2 \sim b^2/(4m^4)$. It gives
a breaking pattern $USp(6) \to \widehat{U(2)}\times USp(2)$; 
(3) as $\La\to 0$, $b,e\to 0$, but $a\neq 0$,
it gives other breaking pattern $USp(6) \to \widehat{U(1)}\times 
USp(4)$. Thus, we see the smooth transition among all of these three
phases.

To count the number of vacua, 
we find the $F_3(t)$ and get the following 
relationship
\bean
-2 m^2= \frac{b^2+3e^2}{2e}+\frac{2\La^4 m^2
(e+m^2)(5e-b)}{e^3(b-e)}.
\nonu
\eean
Note that we use the above conditions $d$ and $e$.  
Combining with the constraint in the above and eliminating $e$, we get 
an equation for $b$ of degree eight. 
Among these eight solutions, three give
a breaking pattern $USp(6) \to USp(0) \times 
\widehat{U(3)}$ which is the same 
as the counting $(N_1-r)=3$, three for $USp(6) \to 
\widehat{U(1)}\times 
USp(4)$ that can be also obtained from the 
counting $(N_0+1)=3$ in the $USp(4)$ part 
and the remainder two, plus the other two from the
$(2,0)$ factorization, for $USp(6) \to \widehat{U(2)}\times USp(2)$
which is equivalent to the counting $(N_1-r) \times (N_0+1)=4$.

Finally we will discuss the following factorization
\bean
t P_3(t)+ 2\La^4 m^4 - 2 \La^4 (t-m^2)^2 & = & 
t (t+a) (t+b)^2, \\
t P_3(t)+ 2\La^4 m^4 + 2 \La^4 (t-m^2)^2 & = & 
(t+c)^2(t+d)^2.
\nonu
\eean
The solution is given by
\bean
u_1 & \equiv & c+d=\frac{3a}{4}+\frac{2(\La^4+\eta \sqrt{2} 
\La^2 m^2)}{a},~~~~ b  =  \frac{a^2+8\La^4+8\eta \sqrt{2} 
\La^2 m^2}{4a}, \\
 u_2& \equiv & c d =\eta \sqrt{2} \La^2 m^2,
-\frac{a^3}{16}-4\La^4 m^2-\frac{4\La^6 (\La^2+\eta \sqrt{2} 
m^2)}{a}+a \left(-\La^4 +\frac{\eta \La^2 m^2}
{\sqrt{2}}\right)=0.
\nonu
\eean
There are eight solutions. However putting back to  curve, we 
find only four different curves. These solutions give only 
one semiclassical limit $\La \to $ with  $a/\La \to 
\mbox{const}$.  
It gives $USp(6)\to USp(6)$. Thus, the four vacua match the 
prediction by $(N+1)=4$ in the weak coupling analysis.

$\bullet$ {\bf Baryonic and Non-baryonic $r=1$ branches}

Next we discuss the $r=1$ branch. There are non-baryonic 
branches for $\widehat{U(3)}$ and $\widehat{U(2)}$. 
For $\widehat{U(1)}$ and  $\widehat{U(3)}$, there are  
baryonic branches. 

By using the addition map, we can reduce the discussion to the one 
for $USp(4)$ gauge theory with $N_f=2, r=0$. 
To better  
understand the phase structure without explicit calculations. 
The first kind of solution gives rise to the breaking pattern 
$USp(6)\to USp(0) \times \widehat{U(3)}$. There is one vacuum. 
The second kind of solution gives rise to the two vacua with 
$USp(6)\to \widehat{U(2)}\times USp(2)$ and one vacuum 
with $USp(0) \times 
\widehat{U(3)}$. These vacua are smoothly interpolated 
to each other. The number of vacua matches the counting from
the weak coupling analysis. Note that we do not obtain the 
breaking pattern $USp(6)\to \widehat{U(1)}\times USp(4)$ because
the number of vacua is zero, 
which matches the counting from 
the weak coupling analysis because of $(N_i-r)=0$.

Finally from the factorization for the baryonic branch, we obtain 
one vacuum with $USp(6)\to USp(0) \times \widehat{U(3)}$ and three vacua 
with $USp(6)\to \widehat{U(1)}\times USp(4)$. These numbers of 
vacua match those in the 
weak coupling analysis. Due to  the baryonic 
branch $\widehat{U(1)}$ and $\widehat{U(3)}$, we have only one 
vacuum respectively and for the $USp(2N)$ factor there exist  
$(N+1)$ vacua.

$\bullet$ {\bf Baryonic and Non-baryonic $r=2$ branches}

Next we will discuss the $r=2$ branch. As in the $r=1$ branch, 
we can use the addition map and thus reduce to the discussion for 
$USp(4)$ theory with $r=1, N_f=2$. From this discussion, we 
can obtain two kinds of solutions. The first kind of solution, which 
comes from $r=1$ branch for $USp(4)$ theory, gives rise 
to the two vacua with the breaking pattern 
$USp(6)\to USp(0) \times \widehat{U(3)}$. The other 
kind of solution gives rise to the two vacua with $USp(6)\to 
\widehat{U(2)}\times USp(2)$. These numbers of vacua match 
those from the weak coupling analysis. 
On the baryonic branch, $\widehat{U(2)}$ has one vacuum 
 on the non-baryonic branch and  $\widehat{U(3)}$ has two
vacua because of $(N_i-r)=2$.

\subsubsection{$N_f=6$ }

The curve is given by
\bean
 ty^2=\left[ t (t^3 -s_1 t^2+ s_2 t-s_3)+ 2\La^2 m^6 \right]^2- 
4 \La^{4} (t-m^2)^6.
\nonu
\eean

$\bullet$ {\bf Non-baryonic $r=0$ branch}

Now let us discuss the $r=0$ branch. There are only non-baryonic 
branches for
$\widehat{U(3)}$, $\widehat{U(2)}$, and $\widehat{U(1)}$. For 
the factorization $(2,0)$ it is given by
\bean
t (t^3 -s_1 t^2+ s_2 t-s_3)+ 2\La^2 m^6 + 2 \La^2 (t-m^2)^3 & = & 
t F_3(t), \\
t (t^3 -s_1 t^2+ s_2 t-s_3)+ 2\La^2 m^6 - 2 \La^2 (t-m^2)^3 & = & 
(t^2+a t+b)^2.
\nonu
\eean
The solutions are given by
\bean
s_1=-2(a+\La^2),~s_2=a^2-2\eta \La m^2(3\eta \La +2m),
~s_3=2\eta \La m^3(2a-3\eta \La m),~
b=-2\eta \La m^3
\nonu
\eean
which give the breaking $USp(6) \to \widehat{U(2)}\times  USp(2)$.
Calculating the $F_3(t)$, we get $a=m^2$ with two vacua.

For the
factorization $(1,1)$, the curve is  given by
\bean
t P_3(t)+ 2\La^2 m^6 + 2 \La^2 (t-m^2)^3 & = & 
t (t+a) (t+b)^2, \\
t P_3(t)+ 2\La^2 m^6 - 2 \La^2 (t-m^2)^3 & = & 
(t^2+c t+d)(t+e)^2.
\nonu
\eean
The solution is given by 
\bean
c&=&\frac{b \left( b + e \right) }{2 e} - \frac{2
{\Lambda}^2 m^2
     \left( 3 b e^2 + 6 e^2 m^2 - b m^4 + 4 e m^4 
\right) }{e^3 \left( -b + e \right) }, \\
a&=&\frac{e \left( b + e \right) }{2b} + \frac{6
{\Lambda}^2 m^2 {\left( e + m^2 \right) }^2}
   {b \left( b - e \right) e} ,\ \ d=\frac{4\La^2 m^6}{e^2}, \\
0&=&\frac{-{\left( b - e \right) }^3}{2} - \frac{2
{\Lambda}^2 {\left( e + m^2 \right) }^2 
     \left( 2 b e - b m^2 + 3 e m^2 \right) }{e^2}.
\nonu
\eean
There are the following limits: (1) when $\Lambda \to 0,b\sim c 
\neq 0$, $(b-e)\sim \Lambda^{2/3}$. It gives 
$\widehat{U(3)}$; (2) when 
$\La \to 0,b\neq 0$, there is $\La^2/e^2\sim b^2/4m^6$ which  
gives $USp(6)\to \widehat{U(2)}\times USp(2)$; (3) when $ \La 
\to 0, b,e\to 0$ with $a\neq 0$, it gives a breaking pattern 
$USp(6)\to 
\widehat{U(1)}\times USp(4)$. Thus, we see a smooth 
transition among all of these three phases.

To count the number of 
vacua, we find the $F_3(t)$ and get the following 
relationship
\bean
-2 m^2= \frac{\left( b + e \right) \left( b^2 + 
e^2 \right) }{2 b e} + 
  \frac{2 {\La}^2 m^2 \left( 3 e^2 {\left( e + 
m^2 \right) }^2 + 
       2 b e m^2 \left( 3 e + 2 m^2 \right)  + 
b^2 \left( 3 e^2 - m^4 \right)  \right) }{b
     \left( b - e \right)  e^3}.
\nonu
\eean
Combining with the constraint in the above and eliminating $e$,
 we get 
an equation for $b$ of degree eight. Among these eight solutions, 
three of them give
$USp(6) \to USp(0) \times \widehat{U(3)}$, three give $USp(6) \to 
\widehat{U(1)}\times 
USp(4)$ and the remainder two, plus the other two from the
$(2,0)$ factorization, for $USp(6) \to \widehat{U(2)}\times USp(2)$.
The consistency with the weak coupling analysis is similar to 
the $r=0, N_f=4$ case.

Finally let us discuss the following factorization
\bean
t P_3(t)+ 2\La^2 m^6 + 2 \La^2 (t-m^2)^3 & = & 
t (t+d-4\La^2) (t+c)^2, \\
t P_3(t)+ 2\La^2 m^6 - 2 \La^2 (t-m^2)^3 & = & 
(t^2+at+b)^2.
\nonu
\eean
 The solution gives
\bean
a&=&\frac{3 d^2 - 32 d {\Lambda}^2 - 8 \Lambda
     \left( -8 {\Lambda}^3 + 6 \Lambda m^2 + 2 \eta m^3 
\right) }{4 d},\\
c&=&\frac{d^2 - 16 d {\Lambda}^2 - 8 \Lambda
     \left( -8 {\Lambda}^3 + 6 \Lambda m^2 + 2 \eta
m^3 \right) }{4 d}.
\nonu
\eean
It is easy to solve numerically.  
There exist ten solutions. However, six of 
them have $s_1\to m$ and the remaining of them 
have $s_1\to 0$, which is the one we are 
looking for and which gives $USp(6)\to USp(6)$. 
The counting of vacua in the weak coupling analysis 
gives $(N+1)=4$, which is the same as  in the above 
strong coupling analysis.

$\bullet$ {\bf Baryonic and Non-baryonic $r=1$ branches}

Next we will  discuss the $r=1$ branch. There are non-baryonic 
branches for $\widehat{U(3)}$ and $\widehat{U(2)}$. 
For $\widehat{U(1)}$, there is also the baryonic branch. 

By using the addition map, we can reduce the discussion to the one 
for $USp(4)$ theory with $N_f=4, r=0$. We can 
understand the phase structure without explicit calculations. 
The first kind of solution for $USp(4)$ case gives rise to one 
vacuum with a breaking pattern 
$USp(6)\to USp(0) \times \widehat{U(3)}$. The second 
kind of solution gives rise to the two vacua with $USp(6)\to 
\widehat{U(2)}\times USp(2)$ and one vacuum with $USp(6)\to 
USp(0) \times \widehat{U(3)}$. These vacua are smoothly interpolated to 
each other. The number of vacua matches the counting from 
the weak coupling analysis. As in $N_f=4$ case, we do not obtain 
the breaking pattern $USp(6)\to \widehat{U(1)}\times USp(4)$, 
which matches the one from 
the weak coupling analysis because  $(N_i-r)=0$.

Finally from the factorization for baryonic branch, we obtain 
three vacua with $USp(6)\to \widehat{U(1)}\times USp(4)$. 
Thus, the number of vacua matches with the prediction by $(N+1)=3$. 

$\bullet$ {\bf Non-baryonic  $r=2$ branch}

Next let us discuss the $r=2$ branch. As in the $r=1$ branch, we 
can use the addition map and reduce to the discussion for $USp(4)$ 
with $r=1, N_f=4$. We obtain only 
one vacuum with $USp(6)\to \widehat{U(3)}$. This number of 
vacuum matches the prediction    from the weak coupling
analysis  that $(N-r)=1$

\begin{table} 
\begin{center}
\begin{tabular}{|c|c|c|c|c|c|c|} \hline
 $N_f$  &  Group  & Branch &  Power of $t(=x^2)$ &  $U(1)$ & 
Number of vacua & Connection
 \\ \hline
 0 & ${USp(6)}$  & $(C)$ &  $t^1$ & 0 & 4 & \\ \cline{2-7}
 & $USp(2) \times {U(2)}$  & $(C,0_{NB})$ &  $t^1$ & 1 & 2 &  \\ \cline{3-7}
 &  & $(C,0_{NB})$ &  $t^1$ & 1 & 2 & A \\ \cline{2-7}
& $USp(4)\times U(1)$ & $(C,0_{NB})$ & $t^1$ & 1 & 3 & A \\ \cline{2-7}
& $USp(0) \times U(3)$ & $(C,0_{NB})$ & $t^1$ & 1 & 3 & A \\ \hline 
2 & ${USp(6)}$  & $(C)$ &  $t^1$ & 0 & 4 & D \\ \cline{2-7}
 & ${USp(4)}\times \widehat{U(1)}$  & $(C,1_{B})$ &  $t^1$ & 0 & 3 & \\ \cline{3-7}
 &   & $(C,0_{NB})$ &  $t^1$ & 1 & 3 & C \\ \cline{2-7}
& $USp(2) \times \widehat{U(2)}$ & $(C,1_{NB})$ & $t^1$ & 1 & 2 & B \\ \cline{3-7}
&  & $(C,0_{NB})$ & $t^1$ & 1 & 2 &  \\ \cline{3-7}
&  & $(C,0_{NB})$ & $t^1$ & 1 & 2 & C \\ \cline{3-7}
&  & $(C,0_{B})$ & $t^1$ & 0 & 2 & D \\ \cline{2-7}
& $USp(0) \times \widehat{U(3)}$  & $(C,1_{NB})$ &  $t^1$ & 1 & 2 &B \\\cline{3-7}
 &   & $(C,0_{NB})$ &  $t^1$ & 1 & 4 &C \\ \hline 
4 & ${USp(6)}$  & $(C)$ &  $t^1$ & 0 & 4 &  \\ \cline{2-7}
 & ${USp(4)}\times \widehat{U(1)}$  & $(C,0_{NB})$ &  $t^1$ & 1 & 3 &F \\ \cline{3-7}
 &   & $(C,1_{B})$ &  $t^1$ & 0 & 3 & G \\ \cline{2-7}
& $USp(2) \times \widehat{U(2)}$ & $(C,0_{NB})$ & $t^1$ & 1 & 2 &  \\ \cline{3-7}
&  & $(C,0_{NB})$ & $t^1$ & 1 & 2 & F \\ \cline{3-7}
&  & $(C,1_{NB})$ & $t^1$ & 1 & 2 & E \\ \cline{3-7}
&  & $(C,2_{B})$ & $t^1$ & 0 & 2 &  \\ \cline{2-7}
& $USp(0) \times \widehat{U(3)}$  & $(C,0_{NB})$ &  $t^1$ & 1 & 3 &F \\\cline{3-7}
&   & $(C,1_{NB})$ &  $t^1$ & 1 & 1 & \\\cline{3-7}
&  & $(C,1_{NB})$ &  $t^1$ & 1 & 1 &E \\\cline{3-7}
&   & $(C,1_{B})$ &  $t^1$ & 0 & 1 &G \\\cline{3-7}
 &   & $(C,2_{NB})$ &  $t^1$ & 1 & 2 & \\ \hline 
6 & ${USp(6)}$  & $(C)$ &  $t^1$ & 0 & 4 &  \\ \cline{2-7}
 & ${USp(4)}\times \widehat{U(1)}$  & $(C,0_{NB})$ &  $t^1$ & 1 & 3 & H \\ \cline{3-7}
 &   & $(C,1_{B})$ &  $t^1$ & 0 & 3 &  \\ \cline{2-7}
& $USp(2) \times \widehat{U(2)}$ & $(C,0_{NB})$ & $t^1$ & 1 & 2 &  \\ \cline{3-7}
&  & $(C,0_{NB})$ & $t^1$ & 1 & 2 & H  \\ \cline{3-7}
&  & $(C,1_{NB})$ & $t^1$ & 1 & 2 & J \\ \cline{2-7}
& $USp(0) \times \widehat{U(3)}$  & $(C,0_{NB})$ &  $t^1$ & 1 & 3 &H \\\cline{3-7}
&   & $(C,1_{NB})$ &  $t^1$ & 1 & 1 &J \\\cline{3-7}
&   & $(C,1_{NB})$ &  $t^1$ & 1 & 1 & \\\cline{3-7}
 &  & $(C,2_{NB})$ &  $t^1$ & 1 & 1 & \\ \hline 
\end{tabular} 
\end{center}
\caption{\sl The summary of the phase structure of 
the $USp(6)$ gauge group with massive flavors. 
In particular, the phases having the smooth transitions
$USp(2N_0) \times U(N_1) 
\leftrightarrow USp(2M_0)\times U(M_1) \leftrightarrow 
USp(2L_0)\times U(L_1) $ 
give the dualities between three different gauge groups (not
two) which were not present in the $SO(6)$ gauge theory with massive 
flavors. As we increase the rank of the gauge group $USp(2N_c)$, 
there will be more dualities connecting more groups.}
\label{massivetableusp6}
\end{table}

\section{Quartic superpotential with massless flavors}
\setcounter{equation}{0}

\indent

In this section, we will consider $USp(2N_c)$ gauge theories 
with $N_f$ {\it massless} flavors. As previously discussed, 
we have studied the massive flavor case and  the tree 
level superpotential had an extremization at $x=\pm m$. However, for 
$USp(2N)$ case in this section,
 since we assume that all the flavors are massless, 
the tree level superpotential is extremized at 
$x=0$. Thus, we have one free parameter which we denote by 
$\alpha$ and as in the $SO(N_c)$ case \cite{afo}
the quartic tree level superpotential is given by 
\begin{eqnarray}
W^{\prime}(x)=x(x^2-\alpha^2). 
\label{spextree}
\end{eqnarray}

In these examples, the gauge group breaks into the two factors, 
$USp(2N_c)\to \widehat{USp(2N_0)}
\times U(N_1)$ where $2N_c=2N_0+2N_1$ and
$N_0\geq 0,~N_1\geq 0$, under 
the semiclassical limit, $\Lambda \to 0$ (Of course, it includes the
degenerated case $USp(2N_c) \to \widehat{USp(2N_c)}$ 
and $USp(2N_c) \to
U(N_c)\times \widehat{USp(0)}$). We consider 
the examples that all the flavors are massless and charged 
under the $\widehat{USp(2N_0)}$. \footnote{As in \cite{bfhn,afo}, 
we use the notation for hat in $\widehat{USp(2N_0)}$ to 
denote a gauge theory with flavors charged under the 
$USp(2N_0)$ group.} The Seiberg-Witten curves for 
these gauge theories are given by 
\begin{eqnarray}
x^2 y^2&=&\left(x^2P_{2N}(x) \right)^2-4(-1)^{N_f}
\Lambda^{4N-2N_f+4}x^{2N_f}.
\nonu
\end{eqnarray}
(or sometimes we use $t=x^2$).

Notice that since the $U(N_1)$ factor does not have any
flavors, it can only be the non-baryonic branch while the
$USp(2N_0)$ factor can be either Chebyshev branch with
 $t^{2[N_f/2]+1}$ or the Special branch with $t^{2(N_f-N_0-1)}$
when $N_f\leq 2N_0+2$ or $t^{2(N_0+1)}$ when $N_f\geq 2N_0+2$. 
It is noteworthy that {\it the power of $t$ is odd 
for the Chebyshev branch and even for the Special branch.}

Now we are ready to deal with the explicit examples for $USp(4)$ and
$USp(6)$ gauge theories.

\subsection{$USp(4)$ case}

\subsubsection{$N_f=1$ }

The curve is given by
\bean
t y^2= \left[ t (t^2 -s_1 t + s_2) \right]^2 + 4 \La^{10} t.
\nonu
\eean

$\bullet$ {\bf Non-degenerated case}

First the curve takes the following factorization form
\bean
t y^2= t F_3(t) H^2_1(t)= t (t^3 +a t^2+ bt+c) (t+d)^2.
\nonu
\eean
The solutions are
\bean
s_1 & = & -2d \pm { \La^5 \over d^{3/2}},~~~s_2=d^2 \mp 
{3\La^5 \over \sqrt{d}}, \\
a & = & 2d\mp  { 2\La^5 \over d^{3/2}},~~~
b= d^2\mp{6\La^5 \over \sqrt{d}}+{\La^{10}\over d^3},~~~~c=
{4 \La^{10}\over d^2}. 
\nonu
\eean
There are two limits:  (1) when $\La\to 0$, but $d\to \mbox{constant}$,
we have $USp(4) \to U(2)\times\widehat{USp(0)} $; 
(2) when $\La\to 0$, $d\to 0$, but 
$\La^5/d^{3\over 2}\to \mbox{constant}$, it gives
 $USp(4) \to U(1)\times  \widehat{USp(2)}$. Thus, we see a
smooth transition from $U(2) \times \widehat{USp(0)}$ 
to $U(1)\times  \widehat{USp(2)}$.
Notice that for $\widehat{USp(2)}$ and $\widehat{USp(0)}$
 with $N_f=1$, there is no
Special branch because the number $(N_f-N-2)$ becomes negative. 
For the Chebyshev branch, there is factor $t^{N_f}=t$, as
we see above. Because of the same factor $t$, this 
explains why we 
observe a smooth transition in this case. 

To count the number of vacua, using the relationship 
\bean
-2 \al^2 =a
\nonu
\eean
we get five solutions totally. Two of them give 
$U(2)\times\widehat{USp(0)}$ where the number of vacua can be
obtained also from the weak coupling analysis $2\times(2N_0+2-N_f)=2$ 
and three others,
$U(1)\times  \widehat{USp(2)}$ which match with the counting
$(2N_0+2-N_f)=3$ in the weak coupling analysis.

$\bullet$ {\bf Degenerated case}

To get the $USp(4) \to \widehat{USp(4)}$, we require 
\bean
t y^2= t (t+a) (t^2 + b t +c)^2.
\nonu
\eean
We found five vacua which match with the counting $(2N+2-N_f)=5$.

\subsubsection{$N_f=2$}
The curve is 
\bean
t y^2= \left[ t (t^2 -s_1 t + s_2) \right]^2 - 4 \La^{8} t^2. 
\eean

$\bullet$ {\bf Non-degenerated case}

The curve takes the following factorization form
\bean
t y^2= t F_3(t) H^2_1(t)= t (t^3 +a t^2+ bt+c) (t+d)^2.
\eean
If $d\neq 0$, the solutions are given by
\bean
s_1 & = & -2d,~~s_2=d^2\mp 2\La^4,~~a=2d,~~b=d^2\mp 4 
\La^4,~~~c=0.
\eean
These solutions give the breaking 
$USp(4) \to U(2)\times \widehat{USp(0)}$. By using 
$-2 \al^2=a$, we get two vacua. Additionally since $c=0$, the curve has
a factor $t^2$, which is exactly the character of $\widehat{USp(0)}$
with $N_f=2$ at the 
Special branch. \footnote{Recall the
condition that $\widetilde{N}= N_f-2-N$ with a factor 
$t^{2( \widetilde{N}+1)}$.} Note also (\ref{example-3}).

If $d=0$, the solution is
\bean
s_2 = \mp 2 \La^4,~~~a=-2 s_1,~~~b=\mp 4\La^4+s_1^2,~~~~c= \pm 4 \La^4 
s_1
\eean
which gives $USp(4) \to U(1)\times  \widehat{USp(2)}$ with
a factor $t^3$. To count the number of 
vacua,  we need to use $-2\al^2=a$ and find that
$s_1=\al^2$. This gives two vacua of 
$USp(4) \to U(1)\times  \widehat{USp(2)}$, which match with the counting
$(2N_0+2-N_f)=2$ in the weak coupling approach.

$\bullet$ {\bf Degenerated case}

\noindent
To get $USp(4) \to \widehat{USp(4)}$, we require 
\bean
t y^2= t (t+a) (t (t+b))^2.
\eean
Solving this equation we get $b=-4 \eta \La^2$, $a= 2\eta \La^2$
where $\eta$ is $4$-th roots of unity. These four vacua match 
 the counting $(2N+2-N_f)=4$ in the weak coupling analysis.

\subsubsection{$N_f=3$}

The curve is given by
\bean
t y^2= \left[ t (t^2 -s_1 t + s_2) \right]^2 + 4 \La^{6} t^3. 
\nonu
\eean

$\bullet$ {\bf Non-degenerated case}

First, the curve takes the following factorization form
\bean
t y^2= t F_3(t) H^2_1(t)= t (t^3 +a t^2+ bt+c) (t+d)^2.
\nonu
\eean
If $d\neq 0$, we have the following solutions
\bean
s_1=-2d \pm {\La^3\over \sqrt{d}},~~s_2= d^2\pm \sqrt{d} \La^3,~~
a=2d\mp {2 \La^3\over \sqrt{d}},~~b={(\sqrt{d^3}\pm 
\La^3)^2\over d},~~
c=0.
\nonu
\eean
There are two classical limits:  (1) when $\La\to 0$, but $d\to 
\mbox{constant}$,
there is $USp(4) \to U(2)\times  \widehat{USp(0)}$; 
(2) when $ \La\to 0$, $d\to 0$, but 
$\La^3/\sqrt{d}\to \mbox{constant}$, it gives
 $USp(4) \to U(1)\times  \widehat{USp(2)}$.  Thus, we see a
smooth transition from $U(2) \times \widehat{USp(0)}$ 
to $U(1)\times  \widehat{USp(2)}$. 
Notice since $c=0$, we have a factor $t^2$ in the curve. 
This indicates that both $\widehat{USp(0)}$ and $\widehat{USp(2)}$
are at the Special branch. 
To count the number of vacua, using $-2 \al^2=a$, 
we find three solutions. Two of
them give $USp(4) \to U(2)\times  \widehat{USp(0)}$ 
which can be obtained also from 
the weak coupling analysis, and the remaining one gives 
$USp(4) \to U(1)\times  \widehat{USp(2)}$ for the Special branch.

If $d=0$, the solution is given by
\bean
s_2=0,~~~~a=-2 s_1,~~~b=s_1^2,~~~c=4 \La^6.
\nonu
\eean
Using $-2\al^2=a$, we find one vacuum for 
$USp(4) \to U(1)\times  \widehat{USp(2)}$ where $\widehat{USp(2)}$
is at the Chebyshev branch by the factor $t^{N_f}=t^3$ and matches with
 the counting from the weak coupling analysis $(2N+2-N_f)=1$.

$\bullet$ {\bf Degenerated case}

Finally we require the curve to be
\bean
t y^2= t (t+a) (t^2 + b t +c)^2.
\nonu
\eean
There are three vacua with $c=0$ of $USp(4) \to \widehat{USp(4)}$
 which match with the counting $(2N+2-N_f)=3$. Notice that
$\widehat{USp(4)}$ is at the Chebyshev branch by the factor $t^3$.

\subsubsection{$N_f=4$}

The curve is 
\bean
t y^2= \left[ t (t^2 -s_1 t + s_2) \right]^2 - 4 \La^{4} t^4.
\eean 

$\bullet$ {\bf Non-degenerated case}

\noindent
The curve takes the following factorization form
\bean
t y^2= t F_3(t) H^2_1(t)= t (t^3 +a t^2+ bt+c) (t+d)^2.
\eean
For $d\neq 0$, the solution is given by
\bean
s_1 = -2 d\pm 2\La^2,~~~s_2= d^2,~~~a=2d \mp 4\La^2,~~~b=d^2,~~~c=0.
\eean
These are  two vacua of  $USp(4) \to U(2)\times  \widehat{USp(0)}$
 with a factor $t^2$.
Notice that to have a smooth transition to 
$USp(4) \to U(1)\times  \widehat{USp(2)}$, since $\widehat{USp(0)}$ 
is at the
Special branch, we require that $\widehat{USp(2)}$  be at the
Special branch. However, the $\widehat{USp(2)}$ at the Special branch
has the factor $t^4$ which explains why we do not find the smooth
transition.

To have $USp(4) \to U(1)\times  \widehat{USp(2)}$ where 
 $\widehat{USp(2)}$ is at the Special branch, the curve should
have a factor $t^4$. Setting $d=0$, we find the solution
\bean
s_2 =0,~~~a=-2 s_1,~~~b=-4 \La^4+s_1^2,~~~~c=0.
\eean
Using the $-2\al^2=a$ we get one vacuum  for 
$USp(4) \to U(1)\times  \widehat{USp(2)}$ where 
 $\widehat{USp(2)}$ is $r=\widetilde{N}=1$ branch 
at the Special branch.
Notice that {\it there is no Chebyshev branch for $\widehat{USp(2)}$ 
with $N_f=4$ just by the simple counting $(2N+2-N_f)=0$.}

$\bullet$ {\bf Degenerated case}

\noindent

Finally to have $USp(4) \to \widehat{USp(4)}$, we require 
\bean
t y^2= t (t+a) (t^2+bt +c)^2.
\eean
There are three solutions. Two of them with $c=b=0$ give a factor
$t^5$. They are the vacua where $\widehat{USp(4)}$ is at the
Chebyshev branch with $(2N+2-N_f)=2$. 
The third one has $a=b=0$ (so with a factor $t^2$)
and gives $\widehat{USp(4)}$ at the Special branch (the dual rank
is $\widetilde{N}=0$). Note also (\ref{example-8}).

\subsubsection{$N_f=5$}

The curve is given by
\bean
t y^2= \left[ t (t^2 -s_1 t + s_2) \right]^2 + 4 \La^{2} t^5. 
\nonu
\eean

$\bullet$ {\bf Non-degenerated case}

First, the curve takes the following factorization form
\bean
 t y^2= t F_3(t) H^2_1(t)= t (t^3 +a t^2+ bt+c) (t+d)^2.
\nonu
\eean
If $d\neq 0$, there are two solutions
\bean
s_1= -2d \pm 3\sqrt{d} \La,~~s_2=d^2\mp d^{3/2} \La,~~a=2(s+2\La^2)
\mp 6\sqrt{d} \La,~~b=d(\sqrt{d}\mp \La)^2,~~c=0.
\nonu
\eean
These give the breaking $USp(4) \to U(2)\times 
\widehat{USp(0)}$. To count the number of
vacua, 
we need to use the relationship between $F_3(t)$ and $W'(x)$. 
However, now it is modified to 
\bea
F_3(t) - 4 \La^2 t^2= t(t-\al^2)^2 +{\cal O}(t) \Rightarrow 
-2 \al^2+ 4 \La^2=a.
\nonu
\eea
From this relationship, 
we  find two vacua for $USp(4) \to U(2)\times 
\widehat{USp(0)}$ with a factor $t^2$ which 
is also consistent with the counting from the weak coupling analysis.

If $d=0$, we get solution
\bean
s_2 =0,~~a=4 \La^2 -2 s_1,~~~b=s_1^2,~~c=0.
\nonu
\eean
The curve has a factor $t^4$ which gives one vacuum of 
 $USp(4) \to U(1)\times  \widehat{USp(2)}$ where 
 $\widehat{USp(2)}$ is at the Special branch. 
This number of vacua can be computed in the weak coupling analysis as
follows: the $\widehat{USp(2)}$ gauge theory with $N_f=5$ on the 
$r=2$ branch has the number $(\widetilde{N}-r+1)=1$. Let us stress
that although the weak coupling analysis seems to tell us that there
are $r$ different branches characterized by  
$0\leq r\leq N_f-2-N_c$, at the strong coupling analysis, 
(i.e., for the 
Seiberg-Witten curve), they  collapse to the same point, the
Special point, in the curve.

$\bullet$ {\bf Degenerated case}

Finally to have $USp(4) \to \widehat{USp(4)}$, we require 
\bean
t y^2= t (t+a) (t^2+bt+c)^2.
\nonu
\eean
There are two solutions. One  is with $a=c=0$ so having the 
factor $t^4$ 
and gives the Special branch of $\widehat{USp(4)}$. The other
is with $c=b=0$ so having a factor $t^5$ and gives the Chebyshev branch
of  $\widehat{USp(4)}$ by counting $(2N_0+2-N_f)=1$.
We have seen that for the $N_f>N_c+2$ case, 
there are two kinds of vacua. The Chebyshev branch 
has $U(N_f)$ flavor symmetry 
and the Special branch has $SO(2N_f)$ flavor symmetry. 
These two kinds of 
solutions each have one vacuum.

\begin{table} 
\begin{center}
\begin{tabular}{|c|c|c|c|c|c|c|} \hline
 $N_f$  &  Group  & Branch &  Power of $t(=x^2)$ &  $U(1)$ & 
Number of vacua & Connection
 \\ \hline
1 & $\widehat{USp(4)}$  & $(C)$ &  $t^1$ & 0 & 5 & \\ \cline{2-7}
 & $\widehat{USp(2)}\times U(1)$  & $(C,0_{NB})$ &  
$t^1$ & 1 & 3 & A \\ \cline{2-7}
& $  \widehat{USp(0)} \times U(2)$ & $(C,0_{NB})$ & $t^1$ & 1 & 2 & A \\ \hline 
2 & $\widehat{USp(4)}$  & $(C)$ &  $t^3$ & 0 & 4 & \\\cline{2-7}
 & $\widehat{USp(2)}\times U(1)$  & $(C,0_{NB})$ &  $t^3$ & 1 & 2 & 
\\\cline{2-7}
& $\widehat{USp(0)} \times U(2)$ & $(S,0_{NB})$ & $t^2$ & 1 & 2 & \\ \hline 
3 & $\widehat{USp(4)}$  & $(C)$ &  $t^3$ & 0 & 3 &  \\\cline{2-7}
 & $\widehat{USp(2)}\times U(1)$  & $(S,0_{NB})$ &  $t^2$ & 1 & 1 
& B \\\cline{3-7}
 &   & $(C,0_{NB})$ &  $t^3$ & 1 & 1 &  \\\cline{2-7}
& $\widehat{USp(0)}\times  U(2)$ & $(S,0_{NB})$ & $t^2$ & 1 & 2 & B \\ \hline    
4 & $\widehat{USp(4)}$  & $(S)$ &  $t^2$ & 0 & 1 & \\ \cline{3-7}
 &   & $(C)$ &  $t^5 $ & 0 & 2 & \\ \cline{2-7}
 & $\widehat{USp(2)}\times U(1)$  & $(S,0_{NB})$ &  $t^4$ & 1 & 1 
& \\\cline{2-7}
& $\widehat{USp(0)}\times  U(2)$ & $(S, 0_{NB})$ & $t^2$ & 1 & 2 & \\ \hline 
5 & $\widehat{USp(4)}$  & $(S)$ &  $t^4$ & 0 & 1 & \\\cline{3-7}
 &   & $(C)$ &  $t^5$ & 0 & 1 & \\\cline{2-7}
 & $\widehat{USp(2)}\times U(1)$  & $(S,0_{NB})$ &  $t^4$ & 1 & 
1 & \\\cline{2-7}
& $\widehat{USp(0)}\times  U(2)$ & $(S,0_{NB})$ & $t^2$ & 1 & 2 & \\ \hline 
\end{tabular} 
\end{center}
\caption{\sl The summary of the phase structure of $USp(4)$ 
gauge group with massless flavors. The flavors are charged under the 
$USp(2N_i)$ factor. We use $S$ for  the Special 
branch for $USp(2N_i)$ factor. All the conventions are the same as  
in previous Tables. It is also worth  comparing with 
the results for $USp(4)$ gauge theory with massive flavors in the previous
section. 
For the smooth transition in the Special 
branch, the condition $(N_f-N_0-2)=M_0$ is exactly 
the condition for Seiberg
dual pair between $\widehat{USp(2N_0)}$ and $\widehat{USp(2M_0)}$ 
where 
$N_f=3,N_0=1$ and $M_0=0$. For this reason (the number of flavors are
fixed by Seiberg dual condition above), one cannot see any
smooth transitions when one increases the number of flavors, $N_f=4,5$.
This kind of observation for $SO(6)$ with 
one massless flavor was found also in \cite{afo}.}
\label{masslesstableusp4}
\end{table}

\subsection{$USp(6)$ case}

We will do for even number of flavor cases as we did for massive case.

\subsubsection{ $N_f=2$ }

The curve is given by
\bean
t y^2= \left[ t (t^3 -s_1 t^2 + s_2 t-s_3) \right]^2 - 4 \La^{12} t^2. 
\nonu
\eean

$\bullet$ {\bf Non-degenerated case}

First the curve takes the following factorization form
\bean
t y^2= t F_3(t) H^2_2(t)=t [t(t+d)]^2(t^3-at^2+bt+c).
\nonu
\eean
We find two kinds of solutions. The one kind of 
solution is given by
\bean
s_1 & = & -2d+\frac{4\eta \La^6}{d^2},~~~s_2=d^2-\frac{8\eta 
\La^6}{d},\ s_3=2\eta \La^6, \\
a & = & -2d+\frac{8\eta \La^6}{d^2},~~~
b= d^2-\frac{16\eta \La^6}{d}+\frac{16\La^{12}}{d^4},~~~~c=
-4\La^6+\frac{32\La^{12}}{d^3}.
\nonu
\eean
There are the two limits:  (1) as $\La\to 0$,\  $d\to 
\mbox{constant}$,
we have $USp(6) \to U(2)\times \widehat{USp(2)}$; (2) when 
$\La\to 0$, 
$d\to 0$,\  
$\La^6/d^{2}\to \mbox{constant}$, it gives
 $USp(6) \to U(1)\times  \widehat{USp(4)}$. Thus, we see a
smooth transition between these phases. 

To count the number of vacua, 
using the relationship between $F_3(t)$ and 
$W^{\prime}(x)$, there is a relation
\bean
2 \alpha^2 =-2d+\frac{8\eta \La^6}{d^2}.
\nonu
\eean
Taking into account  $\eta$ from this equation, 
we find  six solutions. 
Four of them give $U(1)\times \widehat{USp(4)}$ which can be seen 
by the counting $(2N_0+2-N_f)=4$ in the weak coupling analysis 
and two others,
$U(2)\times  \widehat{USp(2)}$. 

The other kind of solutions is given by 
\bean
s_1  &=& -2d,~~~s_2=d^2,\ s_3=2\eta \La^6, \\
a & = & -2d,~~~
b= d^2, \ \ c=-4\eta \La^6,
\nonu
\eean
where $\eta $ is $2$-nd roots of unity. These two solutions 
give  a breaking pattern 
$USp(6)\to {U(2)}\times \widehat{USp(2)}$. Therefore,  there
exist four vacua in the strong coupling analysis, which is
consistent with the counting of $N_1 \times (2N_0+2-N_f)=4$ in 
the weak coupling analysis. 

$\bullet$ {\bf  Degenerated case}

Next we shall consider the degenerate case:
\begin{eqnarray}
ty^2=H_3^2F_2(t)=(t^3-a t^2+ b t+c)^2(t^2+dt+f).
\nonu
\end{eqnarray}
We find two kinds of solutions. One is given by
\begin{eqnarray}
  d=-a,\ \  c=0,\  f=\frac{a^2}{4}-4 \epsilon \La^4,\ \ 
b=\frac{a^2}{4}-\epsilon \La^4
\nonu
\end{eqnarray}
where $\epsilon$ is $3$-rd roots of unity. 
These solutions give $USp(6)\to {U(3)}$.
In this case also the condition 
$\widetilde{N}=0$ with a factor $t^{2(\widetilde{N}+1)}=t^2$ 
is applied and as we have seen before the correct symmetry breaking 
pattern here is $USp(6) \to \widehat{USp(0)} \times U(3)$. 
The number of vacua in the weak coupling analysis for pure $U(3)$
theory has three. 

The other solutions are 
\bean
a & =& 4\epsilon \La^2,~~~
b= 3\epsilon^2 \La^4,\ \ c=0 ,\ \ d=-4\epsilon \La^2 , f=0 
\nonu
\eean
where $\epsilon$ is $6$-th roots of unity. 
These solutions give $USp(6) \to \widehat{USp(6)}$ with a factor $t^3$. 
The number of solution matches with  the prediction $(2N+2-N_f)=6$
in the weak coupling analysis.

\subsubsection{ $N_f=4$}

The curve is given by
\bean
t y^2= \left[ t (t^3 -s_1 t^2 + s_2 t-s_3) \right]^2 - 
4 \La^{8} t^4. 
\nonu
\eean

$\bullet$ {\bf Non-degenerated case}

First the curve takes following form
\bean
t y^2= t F_3(t) H^2_2(t)=t [(t+e)(t+d)]^2(t^3-at^2+bt+c).
\nonu
\eean
The solution is given by
\bean
 a  =  -2d,~~~b= d^2-4\eta \La^4,~~~~c=0, \ \ e=0,
\nonu
\eean
where $\eta$ is $2$-nd roots of unity. These solutions give 
$USp(6)\to {U(2)}\times  \widehat{USp(2)}$. 
These solutions come from the Special branch of 
$\widehat{USp(2)}$. The number of vacua is also seen from the
weak coupling analysis: Pure $U(2)$ gauge theory sets the number of
vacua at two while the $USp(2)$ has the $r={\widetilde{N}}=1$ 
branch at the 
Special branch. This is consistent with the fact that there is
an overall factor $t^{2(\widetilde{N}+1)}=t^4$ 
in the curve. This breaking pattern 
does not appear in the Chebyshev branch because  $(2N+2-N_f)=0$.

The other solution is given by
\bean
t y^2= t F_3(t) H^2_2(t)=t^2 [(t+a-b)(t+b)]^2(t+c)(t+d)
\nonu
\eean
and 
\begin{eqnarray}
d=\frac{2(a-b)^2}{a} , \ \ c=\frac{2b^2}{a} , \ \          
0=(a-2b)^3+4 a\La^4.
\nonu
\end{eqnarray}
If $a=b$, then $d$ becomes zero. These lead to the unbroken case 
we will discuss below. For the case $a\neq b$, 
there are two limits. (1) $\La \to 0, d\neq 0, b\to 0$. 
It gives $U(1)\times \widehat{USp(4)}$. (2) $ \La \to 0,\  
a,b \neq 0$. It gives $\widehat{USp(0)} \times U(3)$. 

To count the number of vacua, using the relationship 
$F_3(t)$ and $W^{\prime}(x)$ there is 
\begin{eqnarray}
-2\alpha^2=\frac{2b^2}{a}+\frac{2(a-b)^2}{a}. 
\nonu
\end{eqnarray}
Under the  limit of (2), we obtain $2b=a-(a\La^4)^{1/3}, a=-2\al^2$, 
so there are three vacua with $U(3)$ which is also consistent with the 
counting in the weak coupling analysis. On the other hand, for the 
limit (1), we obtain $a=0, \pm 2\La^2$ and $b=0$, 
as we have three vacua. This number of vacua  
matches with the results of  the weak coupling analysis. 
Notice that one of three vacua with $U(1)\times 
\widehat{USp(4)}$ comes from the Special branch of 
$\widehat{USp(4)}$ factor. The remaining two vacua come from 
Chebyshev branch with a factor $t^5$.
%

$\bullet$ {\bf Degenerated case}

Next we will consider degenerate case:
\begin{eqnarray}
ty^2=H_3^2F_2(t)=(t^3-at^2+bt+c)^2(t^2+dt+f).
\nonu
\end{eqnarray}
We find two solutions. One is given by
\begin{eqnarray}
  d=4\eta \La^2,\ \  c=0,\ \ f=0,\ \ b=0
\nonu
\end{eqnarray}
where $\eta$ is $4$-th roots of unity. These solutions 
give $USp(6)\to \widehat{USp(6)}$ with a factor $t^5$. 
The number of vacua 
matches with the prediction $(2N+2-N_f)=4$ in the weak coupling
analysis.

\subsubsection{ $N_f=6$ }

The curve is 
\bean
t y^2= \left[ t (t^3 -s_1 t^2 + s_2 t-s_3) \right]^2 - 
4 \La^{4} t^6. 
\nonu
\eean

$\bullet$ {\bf Non-degenerated case}

The curve takes the following factorization form
\bean
t y^2= t F_3(t) H^2_2(t)=t [(t+e)(t+d)]^2(t^3-at^2+bt+c).
\nonu
\eean
The solution is given by
\bean
s_1  =  -2d+2\eta \La^2,~~s_2=d^2,\ \ s_3=0,\  \ a  = 
 -2d+4\eta \La^2,~~b= d^2,~~c=0, \ \ e=0,
\nonu
\eean
where $\eta$ is $2$-nd roots of unity. These solutions 
give $USp(6)\to {U(2)}\times  \widehat{USp(2)}$. These 
solutions have a $t^4$ factor, and so come from the Special branch 
with $r=\widetilde{N}=3$ branch of $\widehat{USp(2)}$ factor where 
$\widetilde{N}=3$. 
The number of vacua matches  
the prediction of $(\widetilde{N}-r+1)=1$ in the weak coupling 
analysis. 


To get $\widehat{USp(4)}\times U(1)$ with $\widehat{USp(4)}$
 at the Special branch, we need a 
factor $t^6$ so $P_3(t)=t^2 (t-s_1)$.
There is only one  solution. In this case, since 
$\widetilde{N}=2$, the $\widehat{USp(4)}$ has the 
$r=\widetilde{N}=2$ branch at the
Special branch.

To get $\widehat{USp(0)} \times  U(3)$, the curve is factorized as
\bean
t^3 - s_1 t^2 +s_2 t-s_3 - 2\La^2 t^2 & = & (t+a) (t+b)^2, \\
t^3 - s_1 t^2 +s_2 t-s_3 + 2\La^2 t^2 & = & (t+c) (t+d)^2. 
\eean
The solution is given by 
\bean
s_1 & = & \frac{2\left( b^3 - d^3 + 
      \left( b^2 + d^2 \right) {\La}^2 \right) 
    }{-b^2 + d^2},~~~~~
s_2  =  b \left( b + \frac{4 d^2
       \left( -b + d - 2 {\La}^2 \right) }{
         -b^2 + d^2} \right),\\
s_3 & = & \frac{-2 b^2 d^2 \left( b - d + 
      2 {\La}^2 \right) }{b^2 - d^2},~~~~~~
a = \frac{2 d^2 \left( -b + d - 2 {\La}^2 \right)
      }{-b^2 + d^2},\\
c & = & \frac{2 b^2 \left( b - d + 2 {\La}^2 \right) }
  {b^2 - d^2},~~~~~~ 
0 = {\left( b - d \right) }^3 - 8 b d {\La}^2.
\eean
There is only one limit we can take: $b,d \neq 0$, but 
$(b-d)\sim \La^{2/3}$. To count 
the vacua, using $a+c=-2 \al^2$ we have three solutions.

$\bullet$ {\bf Degenerated case}

Next we consider degenerate case:
\begin{eqnarray}
ty^2=H_3^2F_2(t)=t^3 (t^2+at^1+b)^2(t+d).
\nonu
\end{eqnarray}
We find three solutions.
\begin{eqnarray}
d&=&a=0,\ \ b=-\La^4, \nonu \\
d&=&4\eta \La^2,\ \ a=b=0,
\nonu
\end{eqnarray}
where $\eta$ is $2$-nd roots of unity. These solutions give 
$USp(6)\to \widehat{USp(6)}$. The number of vacua matches 
 the prediction $(2N+2-N_f)=2$ in the weak coupling 
analysis which is at the Chebyshev branch with a factor $t^7$ 
and the one comes from 
the Special $r=1$ branch with a factor $t^4$ where the counting 
$(\widetilde{N}-r+1)=(1-r+1)=1$ provides with the $r=1$.

\begin{table} 
\begin{center}
\begin{tabular}{|c|c|c|c|c|c|c|} \hline
 $N_f$  &  Group  & Branch &  Power of $t(=x^2)$ &  $U(1)$ & 
Number of vacua & Connection
 \\ \hline
2 & $\widehat{USp(6)}$  & $(C)$ &  $t^3$ & 0 & 6 & \\ \cline{2-7}
 & $\widehat{USp(2)}\times U(2)$  
& 
$(C,0_{NB})$ &  $t^3$ & 1 & 2 & 
 \\ \cline{3-7}
&   
& $(C,0_{NB})$ &  $t^3$ & 1 & 2 & A \\ \cline{2-7}
& $\widehat{USp(4)}\times U(1)$  & $(C,0_{NB})$ &  $t^3$ & 1 & 4 & A 
\\ \cline{2-7}
& $\widehat{USp(0)} \times U(3)$ & $(S,0_{NB})$ & $t^2$ & 1 & 3 &  \\ \hline 
4 & $\widehat{USp(6)}$  & $(C)$ &  $t^5$ & 0 & 4 & \\ \cline{2-7}
 & $\widehat{USp(2)}\times U(2)$  & $(S,0_{NB})$ &  $t^4$ & 1 & 2 
&  \\ \cline{2-7}
& $\widehat{USp(4)}\times U(1)$  & $(S,0_{NB})$ &  $t^2$ & 1 & 1 &
 B \\ \cline{3-7}
&   & $(C,0_{NB})$ &  $t^5$ & 1 & 2 &
  \\ \cline{2-7}
& $\widehat{USp(0)} \times  U(3)$ & $(S,0_{NB})$ & $t^2$ & 1 & 3 & B \\ \hline 
6 & $\widehat{USp(6)}$  & $(C)$ &  $t^7$ & 0 & 2 &  \\\cline{3-7}
&   & $(S)$ &  $t^4$ & 0 & 1 &  \\\cline{2-7}
 & $\widehat{USp(2)}\times U(2)$  & $(S,0_{NB})$ &  $t^4$ & 1 & 2 &  
\\\cline{2-7}
& $\widehat{USp(4)}\times U(1)$  & $(S,0_{NB})$ &  $t^6$ & 1 & 1 &  
\\\cline{2-7}
& $\widehat{USp(0)} \times  U(3)$ & $(S,0_{NB})$ & $t^2$ & 1 & 3 &  \\ \hline    
\end{tabular} 
\end{center}
\caption{\sl The summary of the phase structure of $USp(6)$ gauge 
group with massless flavors. All the conventions are the same as 
in previous Tables. The phases having smooth transition look
similar to those in 
the $SO(6)$ gauge group with massless flavors where 
$\widehat{SO(N_0)} \times U(N_1) \leftrightarrow \widehat{SO(M_0)} 
\times U(M_1)$,
in the 
context of breaking patterns. For the smooth transition in the Special 
branch, the condition $(N_f-N_0-2)=M_0$ is exactly 
the condition for the Seiberg
dual pair between $\widehat{USp(2N_0)}$ and $\widehat{USp(2M_0)}$ 
where 
$N_f=4,N_0=2$ and $M_0=0$. One can also understand the reason why 
one does not see any smooth transition for $N_f=6$ in Special 
branch by checking the Seiberg dual condition: The number of flavors 
are constrained to this. }
\label{masslesstableusp6}
\end{table}

\section{Discussion}

\indent

In this section, we will discuss  about some observations made recently by
Aganagic et al in \cite{vafa}, which
show
that the `trivial' group $USp(0)$ plays a  role in 
the calculation of
`nontrivial' superpotential and makes an agreement between the matrix
model results and standard gauge theory results for $USp(2N)$ 
gauge theory with an adjoint for all Higgs breaking patterns and for
any $N > 0$ (the F-completion of $USp(2N)$ with an adjoint that differs
from the standard gauge theory UV  completion for only 
$USp(0)$ case),
and by Cachazo in \cite{Freddy} which states that the dynamics 
of $USp(2N)$ is related to the one of $U(2N+2)$ with the confining 
index $t=2$ in the notation of \cite{csw}. 
This observation is related to the mysterious 
behavior of $USp(0)$ in the factorization form described in the current
paper. Here we will show that such a relationship can also be
observed with the Brane setup \cite{Ami,gk}.

Recall that the standard Brane setup 
to get $USp(2N)$ gauge theory
in four dimensions (in type IIA string theory)  
is to put $2N$ D4-branes with 
parallel orientifold plane, $O4$-plane, between the two NS5-branes. 
The orientifold
plane is $O4^+$ (with D4-brane charge $+1$) between two NS5-branes and 
$O4^-$ (with D4-brane charge $-1$) outside these  two NS5-branes. 
If we add
one single D4-brane top on the orientifold plane, the total D4-brane
charge outside the two NS5-branes will be zero due to the 
cancellation of the effect of both single D4-brane  and $O4^{-}$, 
just like the
construction of the $U(N)$ gauge group. However, the total 
D4-brane charge inside the two NS5-branes 
becomes $(2N+2)$. This explains 
the mapping $USp(2N)\to U(2N+2)$. Furthermore, the orientifold 
action hints the $U(2N+2)$ gauge theory side 
to be the confining index $t=2$.

All these brane pictures can also be seen 
from the Seiberg-Witten curve:
\bea
ty^2= \left[ t P_N(t) + 2 \La^{2N+2-M} \prod_{j=1}^M m_j \right]^2
-(-1)^M   4 \La^{2(2N+2-M)}\prod_{j=1}^M (t-m_j^2).
\nonu
\eea
The addition of one single D4-brane top on the orientifold plane can be 
seen from the factors $t$ in front of both $y^2$ and the characteristic 
function $P_N(t)$. The
orientifold action (or the confining index $2$) can be seen by
the factor $t=x^2$. 
Then the whole curve is exactly the same form as the one
of $U(N)$ gauge group.

We can use the Brane setup to discuss the relationship between
$U(N)$ and $SO(N)$ gauge theories. In this case, the orientifold is 
$O4^-$ (with D4-brane charge $-1$) for $SO(2N)$ or 
$\widetilde{O4^-}$ (with D4-brane charge $-1$) for $SO(2N+1)$ between
the two NS5-branes, but $O4^+$/$\widetilde{O4^+}$  (with D4-brane charge
$+1$) outside the two NS5-branes for $SO(2N)/SO(2N+1)$. 
To redeem the D4-brane chargelessness outside
the two NS5-branes so that we can compare with the brane setup of
$U(N)$ gauge group, we need to add some object with negative 
D4-brane charge, i.e., anti-D4-brane, to top on the orientifold
in order to cancel the D4-brane charge for $O4^+$/$\widetilde{O4^+}$.
However, the addition of the anti-D4-brane will break the
supersymmetry completely. Thus, we do not expect the
similar relationship between $U(N)$ and $USp(2N)$ gauge theories we have
described above holds 
between {\it  pure} $SO(N)$ and $U(N)$ in
the supersymmetric gauge theories.

However, if we allow the gauge group to carry the fundamental flavors, 
we may
establish the relationship between $SO(N)$ and $U(N)$ gauge theories. 
The reason is
as follows. To give the flavors in the Brane setup, we just need to add
D4-branes outsides these two NS5-branes. Then we can trade the 
$O4^+$ as one single D4-brane while bringing down one D4-brane inside
these two NS5-branes to make the charge from $O4^-$  be neutral.
After that, we can  guess  that the $SO(2N)$ gauge group with $N_f$ 
flavors has some relationship with $U(2N-1)$ gauge group with $(2N_f+1)$
flavors at the confining index $2$. Obviously, it would be interesting 
to check this picture by Konishi anomaly equation as  in 
\cite{Freddy}.

\vspace{1cm}
\centerline{\bf Acknowledgments}

This research of CA was supported by Korea Research Foundation
Grant(KRF-2002-015-CS0006). CA thanks Dept. of Physics, Tokyo 
Institute 
of Technology where part of this work was undertaken. YO would 
like to thank Hiroaki Kanno for useful discussions. CA and YO thank
Katsushi Ito and Kenichi Konishi for discussions. BF would
like to thank Freddy Cachazo and Oleg Lunin for perfectly 
delightful discussions.
The work of BF is supported by the NSF grant  PHY-0070928.

\appendix

\renewcommand{\thesection}{\large \bf \mbox{Appendix~}\Alph{section}}
\renewcommand{\theequation}{\Alph{section}\mbox{.}\arabic{equation}}

\section{\large \bf Strong gauge coupling approach: superpotential and 
a generalized Konishi anomaly equation for $USp(2N_c)$ case }
\setcounter{equation}{0}

\indent

The superpotential considered as a small perturbation is given by
(\ref{treesup}) and the adjoint scalar chiral superfield $\Phi$
has the form of 
\bea 
\Phi= \left( {1 \atop 0 }{ 0 \atop -1 } \right) \otimes
\mbox{diag} ( i \phi_1, i \phi_2, \cdots,  i \phi_{N_c-r}, 0, 0, 
\cdots, 0).
\nonu
\eea
Let us consider 
a special $(N_c-r)$ dimensional submanifold of the Coulomb branch 
where some of the branching points of the moduli space collide.
On the $r$-th branch, the effective theory is
$USp(2r) \times U(1)^{N_c-r}$ with $N_f$ massless flavors.
Following the previous work \cite{afo}, the exact effective superpotential
near a point with $(N_c-r-n)$ massless monopoles is given by 
\bea 
W_{eff}=\sqrt{2} \sum_{l=1}^{N_c-n-r} M_{l}(u_{2s}) q_l
\widetilde{q}_l + \sum_{s=1}^{k+1} g_{2s} u_{2s}. 
\nonu \eea
By varying this with respect to $u_{2s}$,
we get an equation of motion similar to a pure Yang-Mills theory
except that the extra terms on the left hand side since 
the $u_{2s}$ with $2s > 2(N_c-r)$ are dependent on $u_{2s}$ with 
$2s \leq 2(N_c-r)$. 
There exist $(N_c-n-r)$ equations 
for the $(k+1)$ parameters $g_{2s}$. 

Let us consider a singular point in the moduli space 
where $(N_c-r-n)$ monopoles are massless.
The ${\cal N}=2$ curve of genus $(2N_c-2r-1)$ 
degenerates to a curve of genus $2n$  
and it is given by
\begin{eqnarray}
y^2 
 & = & \left( x^2 P_{(N_c-r)}(x)  \right)^2-
4\Lambda^{2(N_c-r)-(N_f-2r)}x^{(N_f-2r)} \nonu \\
& = & H_{N_c-n-r}^2(x)F_{2(2n+1)}(x),
\nonu
\eea
where the double root part and single root part can be characterized by
the following two even functions respectively
\bea
H_{2(N_c-n-r)}(x) =\prod_{i=1}^{N_c-n-r} (x^2-p_i^2), 
\qquad F_{2(2n+1)}(x)=
\prod_{i=1}^{2n+1} (x^2 -q_i^2)
\nonu
\eea
which is similar to the one in $SO(N_c)$ case.
Moreover, the characteristic function can be written as
$P_{(N_c-r)}(x)=\prod_{I=1}^{N_c-r} (x^2 -\phi_I^2)$.

%

$\bullet$ {\bf Field theory analysis for superpotential}

Now we extend the discussion in \cite{bfhn,afo} to $USp(2N_c)$ gauge theory 
and later generalize to the more general cases. 
The massless monopole constraint 
for $USp(2N_c)$ gauge theory with $N_f$ 
flavors is described 
as follows \footnote{It is straightforward to apply it to 
other cases we have 
developed before where the 
structures of single and double roots are different from the one we
are considering here.}:
\begin{eqnarray}
y^2&=&B_{2N_c+2}^2(x)-4\Lambda^{4N_c-4-2N_f}A(x)=x^2H_{2N_c-2n}^2(x)
F_{2(2n+1)}(x),  \nonu \\
&=&x^2\prod_{i=1}^{l} \left(x^2-p_i^2 \right)^2F_{2(2n+1)}(x),
\quad A(x)\equiv  \prod_{j=1}^{N_f}(x^2-m_j^2)
\label{masslessSp}
\end{eqnarray}
where we used $l$ as the number of massless monopoles
and 
\bea
B_{2N_c+2}(x)=x^2 P_{2N_c}(x) + 2 \La^{2N_c+2-N_f} m^{N_f}.
\nonu
\eea 
We have an effective superpotential with 
$l$ massless monopole constraints (\ref{masslessSp}),
\begin{eqnarray}
W_{{low}}&=&\sum_{t=1}^{k+1}g_{2t} u_{2t}  
+ \sum_{i=1}^{l} \left[L_i \left(B_{2N_c+2}(p_i)-2\epsilon_i \Lambda^
{2N_c+2-N_f}\sqrt{A(p_i)}\right) \right. \nonu \\
& + & \left. Q_i \frac{\partial }{\partial p_i}
\left( B_{2N_c+2}(p_i)-2\epsilon_i \Lambda^{2N_c+2-N_f}\sqrt{A(p_i)}  \right) 
\right] 
\nonumber
\end{eqnarray}
where $L_i,Q_i$ are Lagrange multipliers and $\epsilon_i=\pm 1$. 
From the equation of motion for $p_i$ and $Q_i$ we obtain the following 
equations,
\begin{eqnarray}
Q_i=0, \qquad \frac{\partial }{\partial p_i}\left( B_{2N_c+2}(p_i)-
2\epsilon_i \Lambda^{2N_c+2-N_f}\sqrt{A(p_i)}\right)=0.
\nonu
\end{eqnarray}
The variation of $W_{low}$ with respect to $u_{2t}$ leads to 
\begin{eqnarray}
g_{2t}+\sum_{i=1}^{l}L_i\frac{\partial}{\partial u_{2t}}
\left(B_{2N_c+2}(p_i)-2\epsilon_i \Lambda^{2N_c+2-N_f}\sqrt{A(p_i)}\right)=0.
\nonu
\end{eqnarray}
Since $A(p_i)$ is independent of $u_{2t}$, the third term vanishes. 
By using $P_{2N_c}(p_i)=\sum_{j=0}^{N_c}s_{2j}p_i^{2N_c-2j}$ 
we can obtain 
\begin{eqnarray}
g_{2t}=\sum_{i=1}^{l}L_i\frac{\partial}{\partial u_{2t}} P_{2N_c}(p_i)
=
\sum_{i=1}^{l}\sum_{j=0}^{N_c}L_ip_i^{2N_c-2j}s_{2j-2t}. 
\nonu
\end{eqnarray}

With this relation as in \cite{ookouchi}, the following relation
is obtained:
\begin{eqnarray}
W^{\prime}(x)&=
&\sum_{i=1}^{l} \frac{x^3P_{2N_c}(x)}{x^2-p_i^2} L_i-x
\sum_{i=1}^{l}2\epsilon_i L_i \Lambda^{2N_c+2-N_f}\sqrt{A(p_i)} +
{\cal O}(x^{-1}). 
\nonu
\end{eqnarray}
Defining a new polynomial $B_{2(l-1)}$ of order $2(l-1)$ 
as in \cite{civ}, 
\begin{eqnarray}
\sum_{i=1}^{l} \frac{L_i}{x^2-p_i^2} \equiv \frac{B_{2l}(x)}
{x^2H_{2l}(x)}.
\nonu
\end{eqnarray}
Then the deformed superpotential can be expressed as
\bea
W^{\prime}(x) =   \frac{xP_{2N_c}(x) B_{2l}(x)}
{H_{2l}(x)} + {\cal O}(x^{-1}).
\nonu
\eea

Now we can compare  both sides. In particular, for the power behavior of
$x$  it is easy to see that 
the left hand side behaves like $(2n+1)$ while the right hand side 
behaves like
$(2N_c+1-2l)$ except for the factor $B_{2l}(x)$, and therefore the condition
$l=(N_c-n)$ will give rise to the consistency and the polynomial 
$B_{2l}(x)$ becomes a constant.
By using this relation  and substituting the characteristic polynomial 
$P_{2N_c}(x)$ from the monopole constraints (\ref{masslessSp}),
we can obtain the following relation together with the replacement of
the polynomial $H_{2l}(x)$,
\begin{eqnarray}
F_{4n+2}+\frac{4\Lambda^{4N_c-2N_f+4}
\prod_{i=1}^{N_f}(x^2-m_i^2)}{x^2 \prod_{j=1}^{N_c-n}(x^2-p^2_j)^2}=
\frac{1}{g_{2n+2}^2}\left({W^{\prime }_{2n+1}(x)}^2+
{\cal O}(x^{2n})\right). 
\label{spnewmatrix}
\end{eqnarray}
Thus, if $n >-N_f+2N_c+1$, the effect of flavor changes the geometry.
This is a new feature compared with the pure gauge theory without any 
flavors and was used in the specific examples in the previous sections. 

When the breaking pattern is $USp(2N_c) \rightarrow \prod_{i=1}^{n} 
U(N_i)$, the above analysis must be changed. 
The general form was given in \cite{afo}.
Moreover, 
when the origin does not contain the wrapping D5-branes,  as we did for
$SO(N_c)$ gauge theory, the parts relevant to the single roots of SW curve
can be expressed as the superpotential as follows: 
\bea
F_{4n}+\frac{4\Lambda^{4N_c-2N_f+4}
\prod_{i=1}^{N_f}(x^2-m_i^2)}{ \prod_{j=1}^{N_c-n+1}(x^2-p^2_j)^2}=
\frac{1}{g_{2n+2}^2}\left( \left(\frac{W^{\prime}_{2n+1}(x)}{x} 
\right)^2+{\cal O}(x^{2n-2})\right). 
\label{spnewmatrix1}
\nonu
\eea 
Similarly, when if $n >-N_f+2N_c+1$, 
the effect of flavor changes the geometry and this phenomenon  appeared
in the explicit examples.

$\bullet$ {\bf Superpotential of degree $2(k+1)$ less than $2N_c$}

We would like to generalize (\ref{spnewmatrix}) to $2n < 2k$. 
Let us describe the superpotential in the range 
$2n+2 \leq 2k+2 \leq 2N_c$ for $USp(2N_c)$ gauge theory with ${\it massless}$ 
$N_f$ flavors in the $r$-th branch  with the appropriate constraints
by starting from a pure case and modifying the curve
\begin{eqnarray}
W_{eff} & = &
\sum_{s=1}^{k+1}g_{2s}u_{2s}+\sum_{i=0}^{2N_c-2r-2n}
\left[L_i\oint \frac{B_{2(N_c-r)+2}(x)-2\epsilon_i x^{N_f-2r}
\Lambda^{2N_c+2-N_f}}{(x-p_i)}dx \right. \nonu \\
&+ & \left. B_i\oint \frac{B_{2(N_c-r)+2}(x)-2\epsilon_i  
x^{N_f-2r} \Lambda^{2N_c+2-N_f}}
{(x-p_i)^2}dx \right].
\nonu
\end{eqnarray}
For an equal massive case, we simply replace $x^{N_f-2r}
$ in the numerator with $(x^2-m^2)^{N_f/2-r}$ and 
the function $B_{2(N_c-r)+2}(x)$ can be understood as $x^2 P_{2(N_c-r)}(x)+
2 \La^{2N_c-N_f+2} m^{N_f-2r}$.
The $p_i$'s where $i=0, 1,2, \cdots, (2N_c-2r-2n)$ 
are the locations of the double roots of
$y^2 = B_{2(N_c-r)+2}(x)-4 x^{2N_f-4r}
\Lambda^{4N_c+4-2N_f}$. The massless 
monopole points appear in pair $(p_i, -p_i)$ and both 
the function $ B_{2(N_c-r)+2}(x)-2\epsilon_i  
x^{N_f-2r} \Lambda^{2N_c+2-N_f}$ at $x=\pm p_i$ and its
derivative with respect to $x$ at $x=\pm p_i$ become
zero  where
$i=1,2, \cdots, (N_c-r-n)$. 
The total number of constraints is $(N_c-r-n)$ due 
to the fact that the half of the Lagrange multipliers are not independent.
For given constraints, there are only a Lagrange multiplier 
$L_i$ with $(x-p_i)^{-1}$ and a Lagrange multiplier $B_i$ with
$(x-p_i)^{-2}$.

The variation of $W_{eff}$ with respect to 
the Lagrange multiplier $B_i$ will produce 
\begin{eqnarray}
0&=&\oint \frac{B_{2(N_c-r)+2}(x) -2\epsilon_i  x^{N_f-2r}
\Lambda^{2N_c+2-N_f}}
{(x-p_i)^2}dx \nonu \\
& = &
\left( B_{2(N_c-r)+2}(x) -2\epsilon_i  x^{N_f-2r}
\Lambda^{2N_c+2-N_f} \right)^{\prime}
 |_{x=p_i} \nonu \\
& = & \left(x^2P_{2(N_c-r)}^{\prime}(x)+2xP_{2(N_c-r)}(x)  
-\frac{N_f-2r}{x} 2\epsilon_i  x^{N_f-2r}
\Lambda^{2N_c+2-N_f}
\right)|_{x=p_i } \nonu \\
&=&x^2P_{2(N_c-r)}(x)\left(\mbox{Tr} 
\frac{1}{x-\Phi_{cl}}-\frac{N_f-2}{x}
\right)|_{x=p_i}.
\nonu
\end{eqnarray}
Here we used the fact that there exist the 
equations of motion  for the Lagrange multiplier $L_i$ when we replace
$2\epsilon_i  x^{N_f-2r} \Lambda^{2N_c+2-N_f}$ with $x^2 P_{2(N_c-r)}(x)$
at $x=p_i$ and the 
last equality comes from 
$
\mbox{Tr} \frac{1}{x-\Phi_{cl}}+\frac{2r}{x}
=
\sum_{I=1}^{N_c-r} \frac{2x}{x^2-\phi_I^2}=
{P_{2(N_c-r)}^{\prime}}(x)/
P_{2(N_c-r)}(x)$.
For the variation with respect to $p_j$ one obtains
\begin{eqnarray}
0
=2B_j \oint \frac{B_{2(N_c-r)+2}(x) -2\epsilon_i  x^{N_f-2r}
\Lambda^{2N_c+2-N_f}}{(x-p_i)^3}dx
\nonu
\end{eqnarray}
where the equation of motion for the Lagrange multiplier
$B_i$ is used and therefore there is no $L_i$ term.
This expression does not lead to zero because 
the contour integral is not equal to zero. This implies that 
$B_i=0$. Now we turn to the variation of
$W_{eff}$ with respect to $u_{2s}$
\begin{eqnarray}
0=g_{2s}-\sum_{i=0}^{2N_c-2r-2n}\oint \left[\frac{x^2P_{2(N_c-r)}(x)}
{x^{2s}} \right]_{+} \frac{L_i}{x-p_i}dx.
\nonu
\end{eqnarray}

By performing the multiplication with $z^{2s-1}$ 
and summing over $s$, the first derivative of superpotential 
can be written as 
\begin{eqnarray}
W^{\prime}(z)=\sum_{s=1}^{k+1}
g_{2s}z^{2s-1}=\sum_{i=0}^{2N_c-2r-2n}\oint
\sum_{s=1}^{k+1}z^{2s-1}\frac{x^2P_{2(N_c-r)}(x)}{x^{2s}}\frac{L_i}{(x-p_i)}dx.
\nonu
\end{eqnarray}
As before \cite{ao,afo}, after we introduce the new polynomial
$Q(x)$ together with the fact that there exist only half of the 
Lagrange multipliers and the locations of the double roots, 
one gets
\begin{eqnarray}
W^{\prime}(z)=\oint
\sum_{s=1}^{k+1}\frac{z^{2s-1}}{x^{2s}}\frac{Q(x)x^2P_{2(N_c-r)}(x)}
{xH_{2N_c-2n-2r}(x)}dx.
\nonu
\end{eqnarray}
By performing the summation to the infinite sum so as to get 
a closed form, we get
\begin{eqnarray}
W^{\prime}(z)=\oint \sum_{s=1}^{\infty}\frac{z^{2s-1}}{x^{2s}}
\frac{Q_{2k-2n}(x) x^2P_{2(N_c-r)}(x)}{xH_{2N_c-2n-2r}(x)}dx= \oint z
\frac{Q_{2k-2n}(x)x^2P_{2(N_c-r)}(x)}{x(x^2-z^2)H_{2N_c-2n-2r}(x)}dx.
\nonu
\end{eqnarray}
One can use
\begin{eqnarray}
x^2P_{2(N_c-r)}(x)=x\sqrt{F_{2(2n+1)}(x)}H_{2N_c-2n-2r}(x)+{\cal
O}(x^{-2N_c+2N_f-2r-2})
\nonu
\end{eqnarray}
from the monopole constraints. Additionally, for $USp(2N_c)$ case, the second 
terms can not be ignored due to the $N_f$- and $r$-dependent parts.

By substituting this into the above, we get
\begin{eqnarray}
W^{\prime}(z)=\oint z\frac{y_m(x)}{x^2-z^2}dx
+\oint 
{\cal O}\left(x^{-4N_c+2N_f+2k-5} \right)dx
, \qquad y_m^2(x)
=F_{2(2n+1)}(x) Q_{2k-2n}^2(x).
\nonu
\end{eqnarray}
For the condition $-4N_c+2N_f+2k-5 \geq -1$, the second term does
contribute and one can classify the matrix model curve as follows:
\begin{eqnarray}
y_m^2(x)&=&F_{2(2n+1)}(x)Q^2_{2k-2n}(x)= \left\{
				\begin{array}{ll}
				{W_{2k+1}^{\prime}}^2(x)+
{\cal O}\left(x^{2k} \right) & k\ge 2N_c-N_f+2 \\
				{W_{2k+1}^{\prime}}^2(x)+
{\cal O}\left(x^{4N_c-2N_f+4} \right) & k < 2N_c-N_f+2
				\end{array}
				\right. 
		\nonu	\\	&\equiv& 
{W_{2k+1}^{\prime}}^2(x)+f_{2M}(x),\qquad 2M=\mbox{max}(2k,4N_c-2N_f+4).
\label{matrixcurve}
\end{eqnarray}
When $2k=2n$, we
reproduce (\ref{spnewmatrix}) with $Q_0= g_{2n+2}$. 
The second term on the left hand side of (\ref{spnewmatrix}) 
behaves like as
$x^{2N_f-4(N_c-n-1)-2}= x^{2N_f-4N_c+4n+2}$. Depending on whether the power of
this is greater 
than or equal to $2n=2k$, the role of flavor is effective or not.  
When $n=k > -N_f +2N_c-1$, the flavor-dependent part will contribute the 
$W^{\prime}(x)$ \footnote{ When there are no wrapping D5-branes around the 
origin, then the matrix model curve in this case can be obtained similarly and
summarized by
\begin{eqnarray}
y_m^2(x)&=&F_{4n}(x)Q^2_{2k-2n}(x)= \left\{
				\begin{array}{ll}
				\left(\frac{W_{2k+1}^{\prime}(x)}{x} \right)^2+
{\cal O}\left(x^{2k-2} \right) & k\ge 2N_c-N_f+2 \\
				\left( \frac{W_{2k+1}^{\prime}(x)}{x}
 \right)^2+
{\cal O}\left(x^{4N_c-2N_f+2} \right) & k < 2N_c-N_f+2
				\end{array}
				\right. 
		\nonu	\\	&\equiv& 
\left( \frac{W_{2k+1}^{\prime}(x)}{x} \right)^2+
f_{2M}(x),\qquad 2M=\mbox{max}(2k-2,4N_c-2N_f+2).
\nonu
\end{eqnarray}}. 


$\bullet$ {\bf A generalized Konishi anomaly}


Let us assume that the degree of superpotential is less than $2N_c$ 
and consider the following quantity
\begin{eqnarray}
W^{\prime}(\phi_I)=\sum_{i=0}^{2N_c-2r-2n} \oint
\phi_I\frac{x^2P_{2(N_c-r)}(x)}{(x^2-{\phi_I}^2)} \frac{L_i}{(x-p_i)}dx.
\label{Koni2}
\end{eqnarray}
where we put $B_i=0$.
Therefore, by using the above expression
one can write down the following relation 
 \footnote{
One writes down
$
\mbox{Tr} \frac{W^{\prime}(\Phi_{cl})}{z-\Phi_{cl}}=\mbox{Tr}
 \sum_{k=0}^{\infty}z^{-k-1}\Phi_{cl}^kW^{\prime}(\Phi_{cl})=
 \sum_{k=0}^{\infty}z^{-(2i+1)-1}2\sum_{I=1}^{N_c-r}
  \phi_I^{2i+1}W^{\prime}(\phi_I)
=2\sum_{I=1}^{N_c-r }\phi_I
W^{\prime}(\phi_I)\frac{1}{(z^2-\phi_I^2)}   =
\sum_{I=1}^{N_c-r} \frac{2\phi_I^2}{(z^2-\phi_I^2)}
\sum_{i=0}^{2N_c-2r-2n} \oint
\frac{x^2P_{2(N_c-r)}(x)}{(x^2-\phi_I^2)}\frac{L_i}{(x-p_i)}dx$.
From this we change the expression into the linear combination 
of trace part 
$
\frac{2\phi_I^2}{(z^2-\phi_I^2)(x^2-\phi_I^2)}
=\frac{1}{(x^2-z^2)}\left(z\mbox{Tr}\frac{1}{z-\Phi}-
x\mbox{Tr}\frac{1}{x-\Phi} \right)$. 
}
\begin{eqnarray}
\mbox{Tr} \frac{W^{\prime}(\Phi_{cl})}{z-\Phi_{cl}}=\oint
\sum_{i=0}^{2N_c-2r-2n}
\frac{x^2P_{2(N_c-r)}L_i}{(x^2-z^2)(x-p_i)}
\left(z\mbox{Tr}\frac{1}{z-\Phi}-x
\mbox{Tr}\frac{1}{x-\Phi} \right)dx.
\label{Koni4}
\end{eqnarray}
According to the change of contour integration \cite{ao,afo}
\begin{eqnarray}
\oint_{z_{out}}=\oint_{z_{in}}-\oint_{C_z+C_{-z}},
\nonu
\end{eqnarray}
the first term of (\ref{Koni4}) can be written as
\begin{eqnarray}
\left( \mbox{Tr}\frac{1}{z-\Phi_{cl}} \right) \oint_{z_{out}}
\frac{zQ_{2k-2n+2}(x)x^2P_{2(N_c-r)}(x)}{H_{2N_c-2n-2r}(x)x(x^2-z^2)}dx.
\nonu \eea 
Then one obtains the first term of (\ref{Koni4}) in terms of two
parts, 
after change of an integration,
\bea 
&& \left( \mbox{Tr}\frac{1}{z-\Phi_{cl}} \right) 
\left( \oint_{z_{in}}
\frac{zQ_{2k-2n+2}(x)x^2P_{2(N_c-r)}(x)}{H_{2N_c-2n-2r}(x)x(x^2-z^2)}dx+
\oint_{C_{z}+C_{-z}}
\frac{zQ_{2k-2n+2}(x)x^2P_{2(N_c-r)}(x)}{H_{2N_c-2n-2r}(x)x(x^2-z^2)}dx
\right)
\nonu \\
&& = \left( \mbox{Tr}\frac{1}{z-\Phi_{cl}} \right) 
\left( W^{\prime}(z)- \frac{y_m(z)
z^2P_{2(N_c-r)}(z)}{\sqrt{B^2_{2(N_c-r)+2}(z)-4 z^{2N_f-4r} 
\Lambda^{4N_c+4-2N_f}}} 
\right), \label{First}
\end{eqnarray}
where we have  used 
\begin{eqnarray}
H_{2N_c-2n-2r}(z)=\frac{\sqrt{B^2_{2(N_c-r)+2}(z)-4 z^{2N_f-4r} 
\Lambda^{4N_c+4-2N_f}}}{z\sqrt{
F_{2(2n+1)}(z)}},  y_m^2(z)=F_{2(2n+1)}(z)Q_{2k-2n+2}^2(z).
\nonu
\end{eqnarray}

By simple manipulation, 
\footnote{
The second term of
(\ref{Koni4}) becomes
$
-
\oint
\frac{L_ix^2P_{2(N_c-r)}(x)}{(x-p_i)x(x^2-z^2)}
\mbox{Tr}\frac{x}{x-\Phi_{cl}}dx  = -
\frac{L_i
p_i^2P_{2(N_c-r)}(x=p_i)}{p_i^2-z^2} \mbox{Tr}\frac{p_i}{p_i-\Phi_{cl}}$. 
Then from the equation of motion for $L_i$:
$-
\left( N_f-2 \right)\frac{L_i
p_i^2P_{2(N_c-r)}(x=p_i)}{p_i^2-z^2}
=  -
\oint
\frac{\left( N_f - 2 
\right)x^2P_{2(N_c-r)}(x)}{x^2-z^2} \frac{L_i}{x-p_i}dx$.} 
the second term of (\ref{Koni4})
can be reduced to 
\bea 
 -\sum_{i=0}^{2N_c-2r-2n}\oint
\frac{\left( N_f-2r - 2 
\right)x^2P_{2(N_c-r)}(x)}{x^2-z^2} \frac{L_i}{x-p_i}dx
= 
-\left( N_f- 2 \right) \frac{W^{\prime}(z)}{z}. 
\label{Second}
\end{eqnarray}
Therefore we obtain (\ref{Koni4}) by adding the two contributions
(\ref{First}) and (\ref{Second})
\bea
\mbox{Tr} \frac{W^{\prime}(\Phi_{cl})}{z-\Phi_{cl}} & = &
\left( \mbox{Tr}\frac{1}{z-\Phi_{cl}} \right) 
\left( W^{\prime}(z)- \frac{y_m(z)
z^2P_{2(N_c-r)}(z)}{\sqrt{B^2_{2(N_c-r)+2}(z)-4 z^{2N_f-4r} 
\Lambda^{4N_c+4-2N_f}}} 
\right) \nonu \\
&& -\left( N_f- 2 \right) \frac{W^{\prime}(z)}{z} + 
\frac{(N_f-2)}{z} \frac{y_m(z)
z^2P_{2(N_c-r)}(z)}{\sqrt{B^2_{2(N_c-r)+2}(z)-4 z^{2N_f-4r} 
\Lambda^{4N_c+4-2N_f}}}. 
\nonu 
\eea
Recalling that
\bea 
\mbox{Tr}\frac{1}{z-\Phi_{cl}}= \frac{\left(z^2
P_{2(N_c-r)}(z)\right)^{\prime}}{z^2 P_{2(N_c-r)}(z)} -\frac{2}{z}
\nonu 
\eea
and the quantum mechanical form for 
$\left\langle \mbox{Tr}\frac{1}{z-\Phi}
\right\rangle $ we have described before,
we finally obtain the second term of the above expression:
\bea
&& - \left( \mbox{Tr}\frac{1}{z-\Phi_{cl}} \right) \frac{y_m(z) z^2
P_{2(N_c-r)}(z)}{\sqrt{B_{2(N_c-r)+2}^2(z)-4 z^{2N_f-4r} 
\Lambda^{4N_c+4-2N_f}}} \nonu \\
& = & -
\left(\frac{\left(
z^2 P_{2(N_c-r)}(z)\right)^{\prime}}{z^2 P_{2(N_c-r)}(z)}
-\frac{2}{z}\right)
\frac{y_m(z) z^2 P_{2(N_c-r)}(z)}{\sqrt{B_{2(N_c-r)+2}^2(z)-4 z^{2N_f-4r} 
\Lambda^{4N_c+4-2N_f}}} \nonu \\
& = & -\left( \left\langle \mbox{Tr}\frac{1}{z-\Phi}
\right\rangle-\frac{\left( N_f-2 \right)}{z} \right) y_m(z) -
\frac{\left(N_f-2 \right)}{z} \frac{y_m(z) z^2
P_{2(N_c-r)}(z)}{\sqrt{B^2_{2(N_c-r)+2}(z)-4z^{2N_f-4r} 
\Lambda^{4N_c+4-2N_f}}}.
\nonu 
\eea
Using the relation
\bea
\tr \frac{W^{\prime}(\Phi_{cl})-W^{\prime}(z)}{z- \Phi_ {cl}}=
\left\langle \mbox{Tr}
\frac{W^{\prime}(\Phi)-W^{\prime}(z)}{z-\Phi} \right \rangle,
\nonu
\eea
our generalized Konishi anomaly equation can be summarized as follows:
\bea
\left\langle \tr \frac{W^{\prime}(\Phi)}{z-\Phi} \right\rangle &
=& \left( \left\langle \tr \frac{1}{z-\Phi} \right \rangle -
\frac{\left( N_f-2 
\right)}{z} \right) \left[W^{\prime}(z)-y_m(z)\right] 
\label{Konequation}
\eea
which is the generalized Konishi anomaly equation for $USp(2N_c)$
case with flavors. This reproduces the one in a pure adjoint case \cite{ao}
in which $N_f=0$.
The resolvent of the matrix model is related to 
$W^{\prime}(z)-y_m(z)$.

\section{Addition and multiplication maps}
\setcounter{equation}{0}

\indent

In \cite{csw} it was discussed that all the confining vacua 
were constructed from the Coulomb vacua with lower rank 
 gauge groups by using
 the Chebyshev polynomial through the  multiplication
 map. The vacua with a classical gauge group $\prod_{i=1}^n 
U(N_i)$ are transformed into 
the vacua with $\prod_{i=1}^n U(KN_i)$ for the same superpotential
where $K$ is a multiplication index. This multiplication map was extended 
to the $SO(N_c)/USp(2N_c)$ 
gauge theories in \cite{ao}, the $U(N_c)$  
gauge theories with flavors 
in \cite{bfhn}, and $SO(N_c)$ gauge theories with flavors 
\cite{afo}. In addition to this multiplication map, 
another map (called the addition map) was introduced in \cite{bfhn}.  
Thus, we will also study these two maps, the 
addition map and 
multiplication map for the $USp(2N_c)$ gauge theories with flavors.

$\bullet$ {\bf Addition map}

Starting with $USp(2N)$ and $K$ flavors, 
the curve is characterized by
\bean
t y_{\mbox{old}}^2 = \left[ t P_N(t) +2 
\La_{\mbox{old}}^{2N+2-K} m^K \right]^2 
-(-1)^K 4\La_{\mbox{old}}^{2(2N+2-K)} (t-m^2)^K
\nonu
\eean
where the characteristic function is given by 
\bean
P_N(t)=\sum_{j=0}^N  s_j t^{N-j},~~~~s_0=1.
\nonu
\eean
Now we want to  factorize out $(t-m^2)^{2r}$ with $2r\leq K$, so we
should have 
\bean
 t P_N(t) +2 \La_{\mbox{old}}^{2N+2-K} m^K = 
(t-m^2)^r \widehat{P}_{N+1-r}(t).
\nonu
\eean
Parameterized by
\bean
\widehat{P}_{N+1-r}(t)= \sum_{j=0}^{N+1-r}  \widehat{s}_j t^{N+1-r-j},
\nonu
\eean
we get the following relationship, by writing  $(t-m^2)^r $ in terms of 
binomial expansion,
\bean
  \sum_{j=0}^{N}  s_j t^{N+1-j}  +2 \La_{\mbox{old}}^{2N+2-K} m^K
 =  \left(\sum_{l=0}^r t^{r-l} (-1)^l C^l_r m^{2l} \right) 
\left( \sum_{j=0}^{N+1-r}  \widehat{s}_j t^{N+1-r-j} \right). 
\nonu
\eean
By reading off  the $t$-independent part in both 
sides and recombining the power of $\La$,
the key part is that 
\bean
\widehat{s}_{N+1-r} (-1)^r m^{2r}= 2 \La_{\mbox{old}}^{2N+2-K} 
m^K \Longrightarrow
\widehat{s}_{N+1-r}= 2(-1)^r \La_{\mbox{old}}^{2(N-r)+2-
(K-2r)} m^{K-2r}.
\eean
Now after factorizing out the $(t-m^2)^{2r}$,  by substituting 
the above value for 
$\widehat{s}_{N+1-r}$, the reduced curve 
becomes
\bean
 & & \widehat{P}^2_{N+1-r}(t) 
-(-1)^K 4\La_{\mbox{old}}^{2(2N+2-K)} (t-m^2)^{K-2r} \nonu \\
&& = \left[\sum_{j=0}^{N-r}  \widehat{s}_j t^{N+1-r-j}+
\widehat{s}_{N+1-r} \right]^2
-(-1)^K 4\La_{\mbox{old}}^{2(2N+2-K)} (t-m^2)^{K-2r} \nonu \\
& & = \left[ t\sum_{j=0}^{N-r} \widehat{s}_j t^{N-r-j}+
2 \La_{\mbox{new}}^{2(N-r)+2-(K-2r)} m^{K-2r} \right]^2
\nonu \\
&& - (-1)^{K-2r} 4\La_{\mbox{new}}^{2(2(N-r)+2-(K-2r))} (t-m^2)^{K-2r}
\eean
which is the curve of $USp(2N-2r)$ theory with $(K-2r)$ flavors while
the new coupling scale has the following relationship
\bean
 (-1)^r\La^{2N+2-K}_{\mbox{new}}= \La^{2N+2-K}_{\mbox{old}}.
\eean
Thus, we  have the familiar addition map just like the cases
of $U(N)$ and $SO(N)$ gauge theories.
For example,
the phase structure of $USp(6)$ with $N_f=4$
massive flavors on the $r=1$ branch can be read off the one of
$USp(4)$ with $N_f=2$ massive flavors on the $r=0$ branch using 
the addition map.

$\bullet$ {\bf Multiplication map}



Let us assume that the SW curve for $USp(2N_c)$ gauge theory with $2l$ 
flavors and the scale $\La_0$ has the following massless monopole 
constraint 
\bea
\left[ x^2P_{2N_c}(x)+2 \Lambda_0^{(2N_c+2-2l)} m^{2l} \right]^2-
4 \Lambda_0^{2(2N_c+2-2l)}
(x^2-m^2)^{2l}= f_{2p}(x) H^2_{2N_c+2-p}(x)
\label{ori}
\eea
where $p$ is some number
and let us define \footnote{When we put $l=0$ (flavorless case) into 
(\ref{widex}), then it becomes $\widetilde{x} =\frac{x^2 P_{2N_c}(x)}{
2 \eta \La^{2N_c+2}} +1$ which appeared in the pure case \cite{ao}.}
\bea
\widetilde{x}={ x^2P_{2N_c}(x)+2 \eta\Lambda^{(2N_c+2-2l)} m^{2l} \over 
2 \eta\Lambda^{(2N_c+2-2l)} (x^2-m^2)^l}.
\label{widex}
\eea
We claim that the solution is given by
\bea 
\label{Sp-multi}
x^2 P_{K(2N_c+2)-2}(x)+2 \Lambda^{K(2N_c+2-2l)} m^{2Kl}=
2 \left[\eta\Lambda^{(2N_c+2-2l)} (x^2-m^2)^l \right]^K 
{\cal T}_{K}(\widetilde{x})
\eea
where the phase $\eta$ is determined later.
Before we show it is true, let us see  the consistency of the above 
expression. The key part is
that when $x=0$, the equation should also hold. 
For $x=0$, the left hand side of (\ref{Sp-multi}) is
$2 \Lambda^{K(2N_c+2-2l)} m^{2Kl}$. For the right hand side of
(\ref{Sp-multi}), one realizes that $\widetilde{x}=
(-1)^l$ at $x=0$, so by substituting this into the first kind of 
Chebyshev polynomial one gets
${\cal T}_{K}((-1)^l)=(-1)^{Kl}$. To match  both sides, 
we need to have
\bea
\eta^K=1.
\nonu
\eea
This is consistent with the case in pure adjoint. \footnote{Note that
the $\eta$ here corresponds to $\eta^2$ in \cite{ao}.}

Now we can show that it is a 
really true solution. At first we rewrite the following curve
for $USp(2(K N_c+K-1))$ gauge theory with $2Kl$ 
flavors and the scale $\La$: 
\bea
& & \left[x^2 P_{K(2N_c+2)-2}(x)+2 \Lambda^{K(2N_c+2-2l)} m^{2Kl} \right]^2-4
\Lambda^{2K(2N_c+2-2l)} (x^2-m^2)^{2Kl} \nonu \\
&& =  4\Lambda^{2K(2N_c+2-2l)} (x^2-m^2)^{2Kl} \left[ \left(
{x^2 P_{K(2N_c+2)-2}(x)+2 \Lambda^{K(2N_c+2-2l)} m^{2Kl} \over 
2 (\eta\Lambda^{(2N_c+2-2l)} (x^2-m^2)^l)^K} \right)^2-1 
\right]. 
\label{lefthandside} 
\eea
Then we substitute the solution (\ref{Sp-multi}) by squaring of it 
into the left hand side of the above
\bea
&& 4\Lambda^{2K(2N_c+2-2l)} (x^2-m^2)^{2Kl} \left[ {\cal 
T}^2_{K}(\widetilde{x})-1 \right] \nonu \\
&& =  4\Lambda^{2K(2N_c+2-2l)} (x^2-m^2)^{2Kl} {\cal 
U}^2_{K-1}(\widetilde{x})
 \left[\left({ x^2P_{2N_c}(x)+2 \eta\Lambda^{(2N_c+2-2l)} m^{2l} \over 
2 \eta\Lambda^{(2N_c+2-2l)} (x^2-m^2)^l}\right)^2-1 \right] \nonu 
\eea
after applying the property of Chebyshev polynomial \footnote{
We use a useful relation ${\cal T}_K^2(\widetilde{x})-1 = (\widetilde{x}^2-1) 
{\cal U}_{K-1}^2(\widetilde{x})$.} and plugging the 
expression of $\widetilde{x}$ (\ref{widex}). Finally we obtain
the factorization of a gauge group with higher
rank,  
by using the information for the factorization of a gauge group with lower
rank (\ref{ori}), 
\bea
\left[ (\eta\Lambda^{(2N_c+2-2l)} (x^2-m^2)^l)^{K-1} 
 {\cal U}_{K-1}(\widetilde{x}) H_{2N_c+2-p}(x) \right]^2 f_{2p}(x) 
\label{righthandside}
\eea
with the identification 
$\eta\Lambda^{(2N_c+2-2l)}=\Lambda_0^{(2N_c+2-2l)}$.
The new solutions satisfy the relation by putting the 
left hand side of (\ref{lefthandside}) 
to be equal to the equation 
(\ref{righthandside}) 
\bea
&& \left[x^2 P_{K(2N_c+2)-2}(x)+2 \Lambda^{K(2N_c+2-2l)} m^{2Kl} \right]^2-4
\Lambda^{2K(2N_c+2-2l)} (x^2-m^2)^{2Kl}  \nonu \\
&& =  
 H^2_{K(2N_c+2)-p}(x) f_{2p}(x)
\nonu
\eea
where $ H_{K(2N_c+2)-p}(x) \equiv
(\eta\Lambda^{(2N_c+2-2l)} (x^2-m^2)^l)^{K-1} 
 {\cal U}_{K-1}(\widetilde{x}) H_{2N_c+2-p}(x)$.

It is easy to check the degree of this polynomial really gives 
$K(2N_c+2)-p$ by adding the powers of $x$.
Again we see that $USp(2N_c)$ with $2l$ flavors is mapped to 
$USp(2KN_c+2K-2)$ with $2Kl$ flavors. The vacua constructed in this way
for the $USp(2KN_c+2K-2)$ with $2Kl$ flavors have the ${\it same}$ 
superpotential as the vacua of the $USp(2N_c)$ with $2l$ flavors because
they have the common function $f_{2p}(x)$. If we define $2N_c^{\prime}=
2KN_c+2K-2$, we find  $2N_c^{\prime}+2=K(2N_c+2)$, implying
that under the multiplication map, the combination $(2N_c+2)$ has 
simple  multiplication by $K$, as shown in 
the pure adjoint case \cite{ao}.
For example, when $N_c=2$ and $K=2$, the information on the $USp(4)$
gauge theory with flavors will give $USp(2KN_c+2K-2=10)$ theory with 
flavors which we will not discuss  in this paper.
However, in future studies
we will develop the explicit application of this 
multiplication map, as in \cite{ao}.

\end{document}